\documentclass[sigconf, screen, nonacm]{acmart}

\renewcommand\footnotetextcopyrightpermission[1]{}
\AtBeginDocument{%
  }

\begin{document}

\title[Adaptive Gen-AI Guidance in VR]
{Adaptive Gen-AI Guidance in Virtual Reality: A Multimodal Exploration of Engagement in Neapolitan Pizza-Making}

\author{Ka Hei Carrie Lau}
\email{carrie.lau@tum.de}
\affiliation{%
\institution{Chair of Human-Centered
Technologies for Learning, Technical University of Munich}
\city{Munich}
\country{Germany}
}

\author{Sema Sen}
\email{sema.sen@tum.de}
\affiliation{%
\institution{Chair of Human-Centered
Technologies for Learning, Technical University of Munich}
\city{Munich}
\country{Germany}
}

\author{Philipp Stark}
\email{philipp.stark@keg.lu.se}
\affiliation{%
\institution{Department of Human Geography, Lund University}
\city{Lund}
\country{Sweden}
}

\author{Efe Bozkir}
\email{efe.bozkir@tum.de}
\affiliation{%
\institution{Chair of Human-Centered
Technologies for Learning, Technical University of Munich}
\city{Munich}
\country{Germany}}

\author{Enkelejda Kasneci}
\email{enkelejda.kasneci@tum.de}
\affiliation{%
\institution{Chair of Human-Centered
Technologies for Learning, Technical University of Munich}
\city{Munich}
\country{Germany}
}

\renewcommand{\shortauthors}{Lau et al.}

\begin{abstract}
Virtual reality (VR) offers promising opportunities for procedural learning, particularly in preserving intangible cultural heritage. Advances in generative artificial intelligence (Gen-AI) further enrich these experiences by enabling adaptive learning pathways. However, evaluating such adaptive systems using traditional temporal metrics remains challenging due to the inherent variability in Gen-AI response times. To address this, our study employs multimodal behavioural metrics, including visual attention, physical exploratory behaviour, and verbal interaction, to assess user engagement in an adaptive VR environment. In a controlled experiment with (\( n = 54 )\) participants, we compared three levels of adaptivity (high, moderate, and non-adaptive baseline) within a Neapolitan pizza-making VR experience. Results show that moderate adaptivity optimally enhances user engagement, significantly reducing unnecessary exploratory behaviour and increasing focused visual attention on the AI avatar. Our findings suggest that a balanced level of adaptive AI provides the most effective user support, offering practical design recommendations for future adaptive educational technologies.
\end{abstract}





\maketitle

\section{Introduction}

Intangible cultural heritage (ICH) encompasses diverse traditional practices, from oral histories and crafts to culinary rituals that preserve cultural identity~\cite{Ahmad15032006, brulotte2016edible, unesco2024oral, unesco2004yamato, di2020heritage, LviStrauss2012TheCT}. Neapolitan pizza-making, recognised by UNESCO in 2017, exemplifies ICH by embodying the cultural values and embodied skills of Naples, Italy~\cite{UNESCO_Pizza_Inscription, Stazio2021, ceccarini2011pizza}. However, globalisation and the spread of fast-food culture increasingly threaten the transfer of this rich procedural knowledge, particularly among younger generations~\cite{Kim2021, wilk2006fast}.

Procedural learning typically involves skill acquisition through sequential practice and is fundamental to sustaining ICH traditions~\cite{Dagnino2015}. Unlike static instructional methods, procedural heritage constantly evolves and thrives through situated, interactive, and continuous practice~\cite{Hou2022}. Traditional educational methods, however, generally employ uniform instruction, overlooking individual differences in learners' backgrounds, interests, and pacing, thus limiting learner engagement and instructional effectiveness~\cite{Bernacki2018, Walkington2014}.

Adaptive learning systems that integrate virtual reality (VR) with generative AI (Gen-AI) can overcome these limitations by dynamically aligning instructional content with individual learner profiles and real-time interactions~\cite{KABUDI2021100017, KASNECI2023102274}. Real-time adaptive feedback is particularly valuable for procedural tasks like culinary practices, which typically have a low tolerance for errors due to material costs and preparation complexity~\cite{Brako2024}. Nevertheless, evaluating the effectiveness of adaptive AI systems remains challenging due to the inherent variability of AI-generated responses. This variability highlights limitations in traditional human-computer interaction (HCI) assessment methods, which often assume consistent system behaviour and thus struggle to effectively capture the dynamic, individualised interactions characteristic of adaptive AI systems, emphasising the need for more robust evaluation frameworks~\cite{Xu2023, wangjiayin2024}.

`Engagement' in human–agent interaction involves attention, emotional connection, and behavioural coordination~\cite{Glas2015,Renninger2016}. Observable cues such as gaze patterns and head movements have emerged as valuable engagement indicators in immersive settings\allowbreak~\cite{SCHILBACH2015130}. According to Mayer’s Cognitive Theory of Multimedia Learning, learners actively~\textit{select, organise and integrate} information presented through the visual/pictorial and auditory/verbal channels. Modern devices such as eye-tracking and head-movement tracking allow researchers to observe how learners allocate attention within these channels. As adaptive systems grow more complex, traditional single-modality metrics are limited in capturing the nuanced dynamics of engagement. Thus, multimodal behavioural analysis provides deeper insights, particularly crucial for skill-based immersive learning contexts~\cite{zhang2024, lau2024evaluating}.

Motivated by this research gap, we developed \textbf{Neapolitan Pizza VR}\footnote{\url{https://youtu.be/SOwzEgueJro}}, an immersive learning environment focused on the traditional culinary practice of Neapolitan pizza-making. The system includes an adaptive Gen-AI tutor that dynamically tailors instructional guidance based on user demographics collected from a pre-assessment survey and real-time user actions (e.g., ingredient selection). For instance, the tutor can offer additional cultural context or adjust instructional details based on these inputs. Through an exploratory user study, we evaluate user engagement using multimodal metrics as behavioural proxies. Our study aims to answer the following research question~\textbf{(RQ)}:

\begin{quote} \textit{How do different levels of adaptive generative AI guidance (high, moderate, none) affect multimodal user engagement in a VR-based intangible cultural heritage experience?} \end{quote}

We conducted a between-subjects experiment with 54 participants, employing multimodal metrics (dwell time, head movements, and verbal interaction) as behavioural proxies for learner engagement. Our analysis explores how adaptivity influences learner interaction, attention allocation, and exploratory behaviour during procedural learning tasks within immersive educational contexts.

Our contributions are threefold:

\begin{enumerate} 
\item We introduce an adaptive VR-based system specifically designed for procedural ICH learning. Key findings show that adaptive AI guidance increases visual attention toward the tutor and reduces unnecessary exploratory behaviour.

\item We validate multimodal metrics as holistic indicators of learner engagement. These behavioural measures reveal how varying adaptivity levels influence visual attention and exploratory behaviours, suggesting improved attentional focus, with moderate adaptivity yielding the highest overall engagement.

\item We offer practical design guidelines for adaptive AI-driven VR educational tools, emphasising balanced adaptivity to sustain learner engagement and autonomy in culturally rich procedural learning contexts.
\end{enumerate}

\section{Related Work}

\subsection{Virtual Reality for Procedural Knowledge}
VR can effectively support procedural learning, particularly within cultural heritage contexts, due to its immersive and embodied learning capabilities. By bridging passive observation and active participation, VR enables learners to experience cultural practices through realistic, simulated interactions~\cite{Bekele2018, Damala2012, Hou2022}. This approach has shown significant effectiveness in procedural learning contexts such as traditional crafts~\cite{Dong2021, CARROZZINO201182}, folk dances~\cite{Aristidou2021}, and the revitalisation of historical practices~\cite{lau2024evaluating}. Furthermore, VR supports the digital preservation and transmission of ICH, facilitating authentic cultural experiences across generations~\cite{Bekele2018, Kenderdine2014, Hou2022}.

However, current VR cultural heritage applications often deliver standardised content, limiting personalised learner engagement~\cite{Bekele2018, Damala2012}. Museums increasingly advocate transitioning from merely presenting information (\textit{being about something}) to actively addressing learners' unique needs (\textit{being for someone}), emphasising adaptive experiences~\cite{weil2007being}. Nonetheless, adaptive VR guidance specifically for procedural skills, particularly culinary heritage, remains unexplored. We address this gap by employing adaptive Gen-AI to dynamically tailor VR instruction based on learners' actions and demographics, enhancing engagement in VR-based culinary heritage education.

\subsection{Adaptive AI for Procedural Learning}

Adaptive-AI is particularly well-suited for procedural ICH learning, as these practices constantly evolve and rely on situated interactions rather than static instructions~\cite{Hou2022}. Yet, current educational practices often deliver uniform, static instructions, neglecting individual learners' diverse backgrounds or interests. This fixed approach can limit learner engagement and motivation~\cite{Bernacki2018, Walkington2014}. While Engagement, particularly in informal learning contexts, is strongly tied to situational interest, a psychological state triggered by meaningful or personally relevant material~\cite{Hidi01062006, Renninger2016}. 

Adaptive learning addresses these limitations by tailoring instructional experiences based on learners' characteristics, preferences, and real-time behaviours, thus enhancing motivation and learning effectiveness~\cite{Do2024, KASNECI2023102274, Leong2024}. Recent adaptive systems range from simple rule-based methods to sophisticated Gen-AI frameworks dynamically generating personalised content aligned with users' cognitive profiles and interaction patterns~\cite{Raptis2019, guo2024enhancing}.

In cultural heritage contexts, adaptive approaches such as personalised museum guides~\cite{Stock2007} and narrative systems~\cite{Nappi2024} have demonstrated improved visitor engagement and satisfaction~\cite{Not2019}. However, effectively implementing adaptive systems remains challenging, given concerns about fairness, algorithmic biases, and managing the inherent unpredictability of AI-generated responses~\cite{ErRafyg2024}. Balancing adaptivity is crucial, as excessive adaptivity might overwhelm learners, while minimal adaptivity risks disengagement~\cite{Pu2024, Huang2012}.

To address these challenges and inform effective design, our study employs multimodal behavioural metrics to evaluate adaptive Gen-AI-driven VR experiences objectively. By examining how varying adaptivity levels impact engagement, we provide insights that can help preserve and disseminate ICH more effectively through adaptive-AI in VR education.

\subsection{Multimodal Measures of Engagement}

Measuring engagement in interactive environments involves cognitive, emotional, and behavioural components~\cite{Glas2015}. Traditional self-report methods are limited in their ability to capture moment-to-moment dynamics, particularly in VR, where short interventions (e.g., completing a questionnaire) can disrupt immersion and presence, both of which are critical for learning in VR~\cite {Putze2020}. Consequently, multimodal assessment methods integrating physiological and behavioural metrics have gained prominence for offering real-time, non-intrusive alternatives~\cite{YAREMYCH2019103845, DUBOVI2022104495}.

Eye-tracking and head movements reliably indicate attention, cognitive engagement, and exploratory behaviour in immersive or adaptive settings~\cite{Zander2015, Michael2020, Beach20102019, YAREMYCH2019103845, DUBOVI2022104495}. Eye-tracking metrics like dwell time can predict learning outcomes and support adaptive assessments~\cite{Mikhailenko2022, DUBOVI2022104495, Dominguez2017}. Similarly, head orientation effectively captures attention shifts, spatial interactions, and social engagement~\cite{YAREMYCH2019103845, Loomis2008}. Beyond attentional metrics, gaze and head orientation also function as social cues; mutual gaze, reciprocal gaze behaviours, and physical proximity reflect social cognition and emotional engagement with virtual agents~\cite{SCHILBACH2015130, McCall2017}.

However, multimodal engagement evaluations in cultural heritage have focused primarily on tangible experiences (e.g., galleries, architecture)~\cite{Michael2020, LiNa2022}, leaving procedural culinary heritage underexplored. Our study addresses this gap, evaluating adaptive Gen-AI guidance in VR-based Neapolitan pizza-making via multimodal metrics using a Cooking as Inquiry Learning framework.

\subsection{Inquiry-Based Cooking for Procedural ICH}

Cooking-as-Inquiry Learning (CIL) frames food preparation as an exploratory, experimental practice rather than following fixed recipes~\cite{Brady2011}. This approach aligns with procedural ICH, where tacit skills such as judging dough readiness develop through iterative experimentation and reflection~\cite{TRUBEK2017297}. A recent review highlights CIL’s role in eliciting food memories, underscoring its value in preserving cultural knowledge~\cite{LEE20231}.

In a multimodal VR kitchen, an adaptive Gen-AI tutor identifies whether a learner follows traditional or non-traditional culinary styles and provides guided inquiry accordingly (e.g., modifying steps based on ingredient choices). However, the richness of sensory channels (visual, auditory, haptic) can cause cognitive overload, presenting an ``interface dilemma'' that requires careful balancing of feedback salience and usability~\cite{Bieniek2024}. Guided by CIL, we frame Neapolitan pizza-making as an inquiry process and use behavioural metrics to evaluate learner engagement with the Gen-AI tutor.

\section{Method}

In this section, we describe the technical implementation, apparatus, the demographics of our participants, experimental design, procedure, measures, data processing, and analysis. The Institutional Review Board (IRB) of [blind institute] granted approval for this user study, ensuring adherence to ethical research standards.

\subsection{Technical Implementation}

\begin{figure}[t]
\centering
\includegraphics[width=\columnwidth]{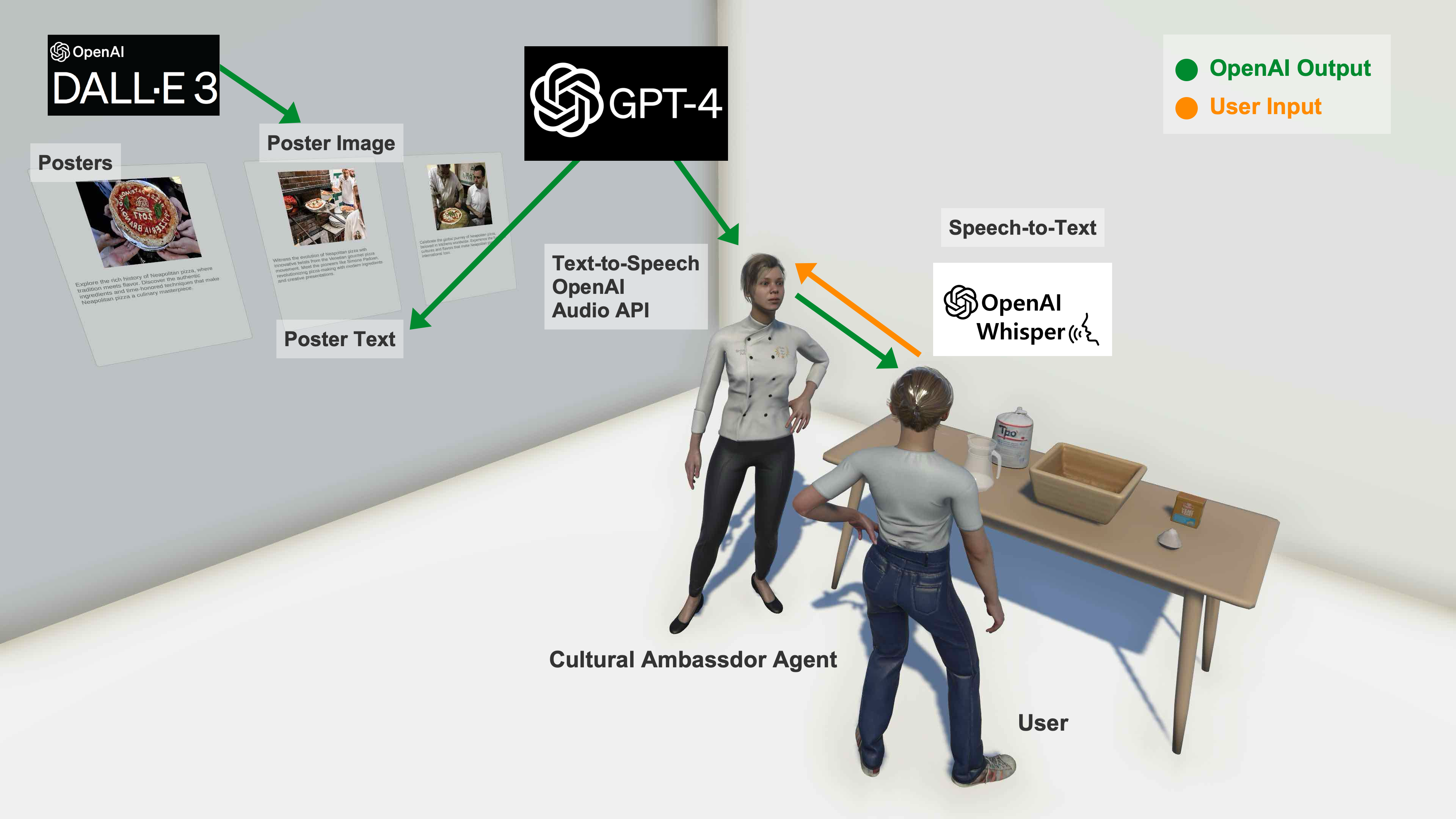}
\caption{Architecture for creating an adaptive VR experience. The simplified VR environment is for illustration only; the actual setup is shown in Figure~\ref{fig:vr_experience}.}
\Description{Architecture for creating an adaptive VR experience. The simplified VR environment is for illustration only; the actual setup is shown in Figure~\ref{fig:vr_experience}.}
\label{fig:api_call}
\end{figure}

The multimodal VR environment \textbf{Neapolitan Pizza VR} was developed around three components: (1) the VR simulation, (2) a pre-prompted virtual coach, and (3) adaptive Gen-AI guidance. Figure~\ref{fig:api_call} shows the overall architecture.

The VR simulation guides participants interactively through traditional pizza-making, from selecting ingredients and preparing dough to baking in a wood-fired oven as shown in Figure~\ref{fig:game_stages}. Object interactions (e.g., grabbing ingredients, kneading dough) utilise Unity’s XR Simple Interactable component\footnote{\url{https://docs.unity3d.com/Packages/com.unity.xr.interaction.toolkit@3.1/manual/xr-simple-interactable.html}, accessed 23 April 2025}. The interactable events trigger handlers that update ingredient states and the adaptive tutor’s instructions.

The virtual coach uses OpenAI's GPT-4 model, pre-prompted to function as a culturally informed instructor. Prompt engineering and few-shot learning ensured culturally appropriate and context-sensitive responses~\cite{sahoo2024systematicsurveypromptengineering, huang2023surveyhallucinationlargelanguage}. Prompts positioned the Gen-AI as a cultural ambassador, defining its instructional style and tone~\cite{mishra-etal-2022-reframing}.

\begin{figure}[ht!]
\centering
\includegraphics[width=0.8\columnwidth]{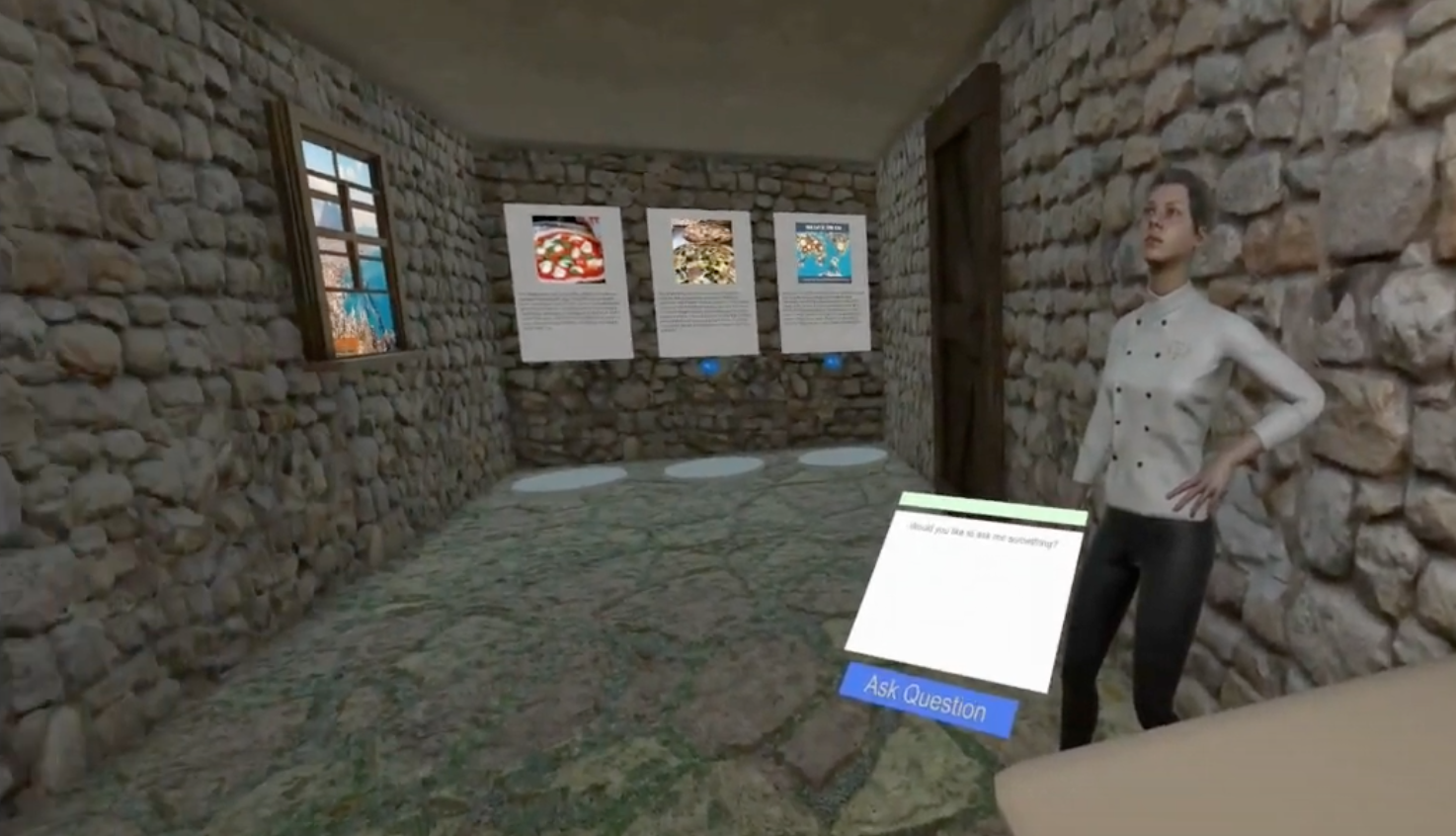}
\caption{VR scene showing avatar interaction within the environment.}
\Description{VR scene showing avatar interaction within the environment.}
\label{fig:vr_experience}
\end{figure}

Adaptive guidance was driven by two inputs: (1) demographic data (age, cooking experience, cultural background) gathered in a pre-assessment as shown in Appendix~\ref{tab:heritage},~\ref{tab:interest},~\ref{tab:culinary} and (2) real-time actions in-VR (i.e., ingredient selection). Choosing non-traditional toppings triggered extra cultural context, while demographics governed explanation depth. GPT-4 generated the verbal guidance and DALL-E produced corresponding visuals, including personalised cultural-history posters. Figure \ref{fig:vr_experience} shows the overall VR environment.

Stage-specific prompts were derived from a Massive Open Online Course (MOOC) on Neapolitan pizza-making\footnote{\url{https://www.federica.eu/federica-pro/pizza-revolution/}, accessed 23 April 2025}. The agent architecture integrated GPT-4, Whisper speech-to-text (STT)\footnote{\url{https://openai.com/index/whisper/}, accessed 23 April 2025}, OpenAI Audio API for text-to-speech (TTS)\footnote{\url{https://platform.openai.com/docs/guides/text-to-speech/quickstart}, accessed 23 April 2025}, and DALL-E. Ingredient models were generated using Luma AI\footnote{\url{https://lumalabs.ai/dream-machine}, accessed 23 April 2025}.

Each STT interaction recorded 5 seconds of participant speech, with GPT-4 responses delivered after a 2–5 second delay (depending on response complexity). Participants reported no noticeable negative impact on their experience.

\subsection{Apparatus}

\begin{figure}[ht!]
\centering
\includegraphics[width=0.8\columnwidth]{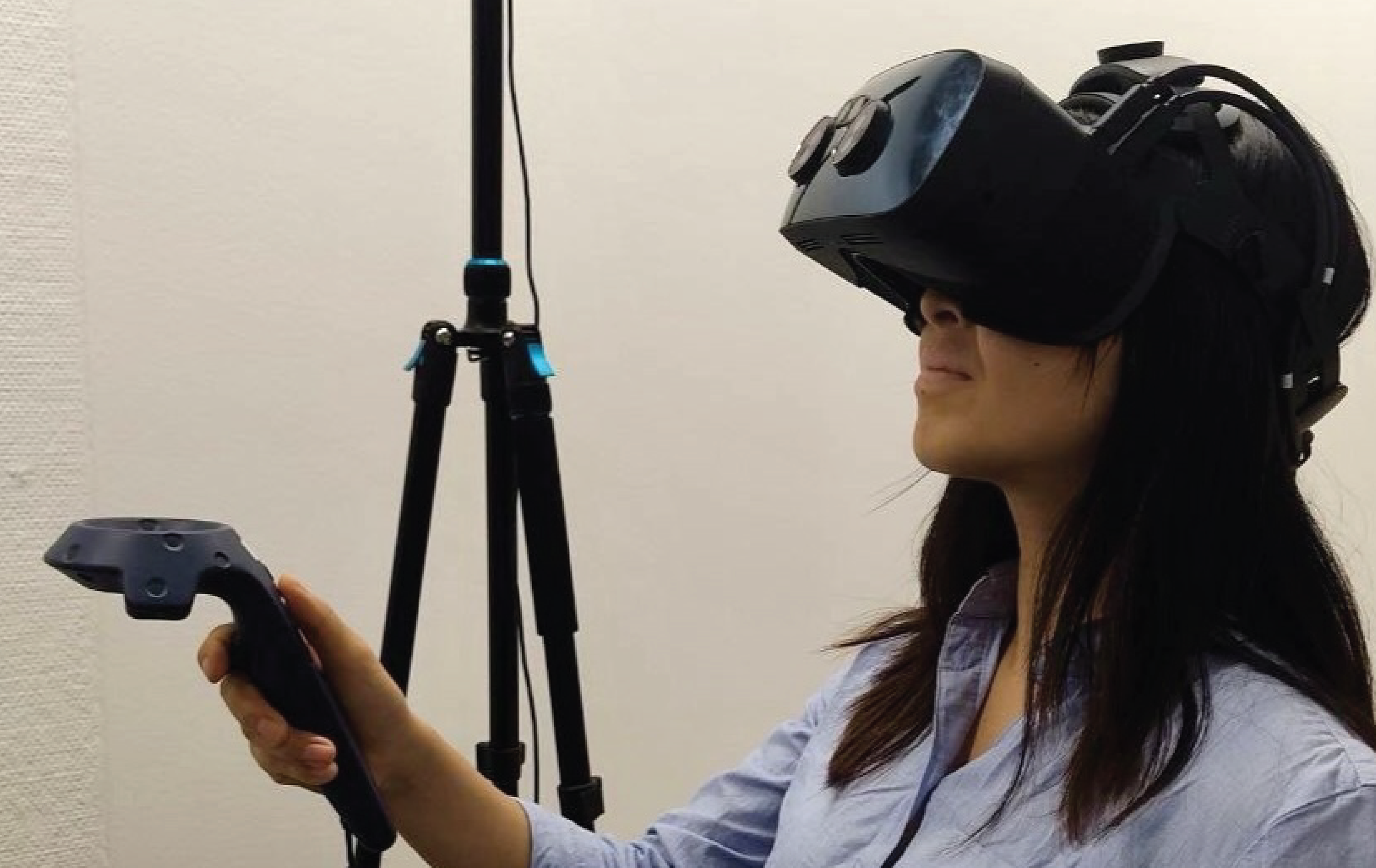}
\caption{Experiment setup with Varjo XR-3 headset and HTC Vive controllers.}
\Description{Experiment setup with Varjo XR-3 headset and HTC Vive controllers.}
\label{fig:VRsetup}
\end{figure}

The VR setup used in this study is shown in Figure~\ref{fig:VRsetup}. It consisted of a Varjo XR-3\footnote{\url{https://varjo.com/products/varjo-xr-3/}, accessed 23 April 2025} headset, paired with HTC Vive Controller 2.0 and HTC Vive Steam VR Base Station 2.0. The Varjo XR-3 offers a 115$^{\circ}$ field of view, a 90 Hz refresh rate, a screen resolution of 1920 × 1920 per eye, and is equipped with a built-in eye tracker that operates at a 200 Hz sampling rate.

The interactive VR game was developed using Unity (Version 2021.3.33f1). Key software extensions included the Varjo XR Plugin\footnote{\url{https://github.com/varjocom/VarjoUnityXRPlugin}, accessed 23 April 2025} (Version 3.6.0) for Varjo-specific support, the XR Interaction Toolkit\footnote{\url{https://docs.unity3d.com/Packages/com.unity.xr.interaction.toolkit@3.0/manual/index.html}, accessed 23 April 2025} (Version 2.5.3), and XR Plugin Management\footnote{\url{https://docs.unity3d.com/2023.2/Documentation/Manual/com.unity.xr.management.html}, accessed 23 April 2025} (Version 4.4.0), both essential for Unity-based VR development. Additionally, the OpenAI Unity\footnote{\url{https://github.com/srcnalt/OpenAI-Unity.git}, accessed 23 April 2025} package (Version 0.2.0) was integrated to enable adaptive AI interactions within the Unity game engine via the OpenAI application programming interface.

\subsection{Participants}

The study involved 54 participants from diverse demographic backgrounds. Gender distribution was nearly balanced with 22 male participants (41\%), 31 female participants (57\%), and one non-binary participant (2\%). Ages ranged from 18 to 54, with the majority in the 18--24 age group (46\%) and the 25--34 age group (48\%). Smaller percentages were in the 35--44 (4\%) and 45--54 (2\%) age brackets.

In terms of educational background, 15\% of participants held a high school diploma, 40\% held a bachelor’s degree, 43\% held a master’s degree, and 2\% held a doctoral degree. Most participants (70\%) had prior experience with VR, while 30\% were new to the technology.

Participants' engagement with cultural heritage activities varied, with 9\% engaging very frequently, 13\% frequently, 41\% rarely, 33\% very rarely, and 4\% reporting no prior engagement. Cooking frequency at home also showed variability: 50\% cooked very frequently, 29.6\% cooked frequently, 18.5\% cooked rarely, and 1.9\% never cooked. These factors were not directly controlled for in this study; future research could explore how these variables influence engagement and learning outcomes.

Eligibility criteria required participants to be at least 18 years old, have normal or corrected-to-normal vision, and be fluent in English. Individuals with a history of severe motion sickness were excluded from the study. Each participant received a \texteuro10 voucher for their involvement at the end of the experiment.

\subsection{Experimental Design}

We employed a between-subjects design with 54 participants randomly assigned to three adaptivity levels: none (baseline), moderate, and high. Participants experienced only one level and were unaware that alternatives existed. In the high condition, the tutor adapted guidance to both demographic data and in-VR actions; in the moderate condition, it adapted only to ingredient choices; the no-adaptivity (baseline) used a fixed script without any adaptation. Example utterances for each level appear in Appendix Table~\ref{tab:tutor_scripts}.

As shown in Figure \ref{fig:vr_experience}, the study took place in a multimodal VR kitchen. Participants wore VR headsets and used controllers to pick ingredients, make the dough, and bake the pizza. STT let participants talk to the tutor, and the tutor always answered via TTS; speaking was optional because the scene itself provided step‑by‑step cues.

Dependent variables as shown in Table~\ref{tab:metrics} were total session time, avatar dwell time, head‑movement speed, and whether the participant spoke to the tutor.

\begin{figure*}[ht]
\centering
\includegraphics[width=0.8\linewidth]{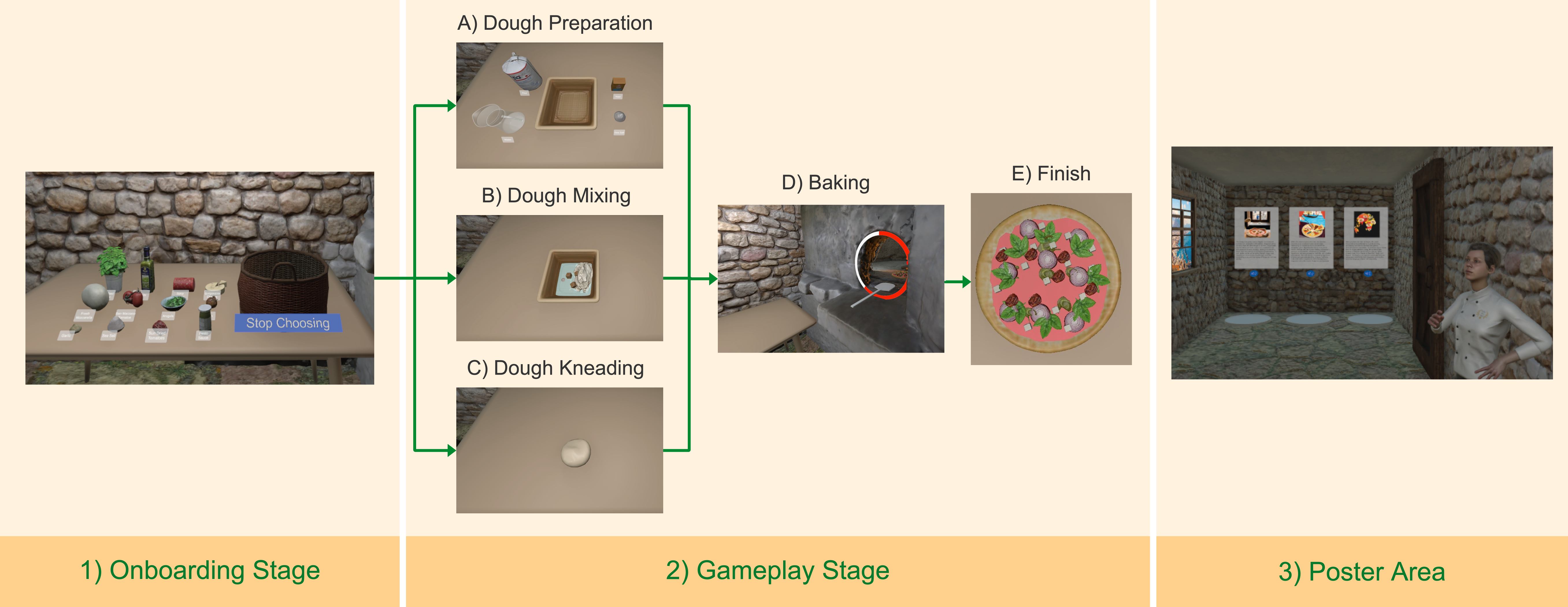}
\caption{Stages in the experimental setup: (1) Onboarding, (2) Gameplay, and (3) Poster exploration. Tutor responses and poster content adapt according to the assigned condition: (non-adaptive baseline), ingredient‑based (moderate), or demographics and behaviour-based (high).}
\Description{Stages in the experimental setup: (1) Onboarding, (2) Gameplay, and (3) Poster exploration. Tutor responses and poster content adapt according to the assigned condition: (non-adaptive baseline), ingredient‑based (moderate), or demographics and behaviour-based (high).}
\label{fig:game_stages}
\end{figure*}

The session proceeded linearly through three stages as shown in Figure~\ref{fig:game_stages}:  
(1) Onboarding, where participants met the virtual agent and chose toppings from twelve ingredients;  
(2) Gameplay, where the tutor guided the traditional pizza‑making steps;  
(3) Poster exploration, where participants examined three cultural‑history posters before the session ended at pizza completion.

\subsection{Procedure}

Prior to the experiment, participants completed a pre-assessment questionnaire to gather baseline data on cultural heritage engagement, cooking preferences, and VR familiarity (in Appendix Tables~\ref{tab:heritage},~\ref{tab:interest}, and~\ref{tab:culinary}). The questionnaire, adapted from~\cite{digital2030020}, included 5-point Likert items to assess motivations for engaging with cultural heritage, such as ``How much do the following reasons motivate you to engage with cultural heritage activities?'' (1 = ``not a motivation,'' 5 = ``very strong motivation''). It also assessed knowledge of Neapolitan pizza and familiarity with VR. Culinary preferences were measured using items from~\cite{Andrew2022} to distinguish traditional and innovative culinary styles. We also collected demographic data, including age, gender, education, and prior VR experience. 

Before the experiment, participants were briefed on the study's purpose, their right to withdraw, and informed of their interaction with a VR pizza-making simulation. They were not told about the Gen-AI agent operating in the background, though they were informed about optional cultural posters within the VR environment.

During the experiment, participants wore a Varjo XR-3 headset while standing. A 5-point eye-tracking calibration was conducted, followed by a few minutes for participants to familiarise themselves with the environment. 

After the task, participants completed a post-assessment, which included questions on their interest in the study, perception of the agent's usability, knowledge of pizza-making steps, and perceived realism of the VR experience. Each session lasted approximately 30-45 minutes, including setup and assessments.

\subsection{Measures}

To evaluate user engagement holistically in adaptive AI-driven VR, we used a multimodal behavioural approach across four dimensions, summarized in Table~\ref{tab:metrics}.

\textbf{Temporal Metric:} Total experiment time (in seconds) served as a baseline measure of sustained engagement~\cite{DUBOVI2022104495, YAREMYCH2019103845}, which may reflect ongoing cognitive-affective involvement in interactive settings~\cite{SCHILBACH2015130}.

\textbf{Visual Attention:} Relative dwell time on the avatar is calculated as the percentage of total session time that participants spent gazing at the AI tutor, serving as an indicator of attentional focus and cognitive engagement~\cite{Bozkir2021, Mikhailenko2022}. Previous research shows that longer dwell times on task-relevant areas predict higher engagement and deeper cognitive processing~\cite{Epstein2022}. Therefore, in our study, higher avatar dwell times are interpreted as reflecting greater cognitive engagement with the adaptive content.

\textbf{Physical Exploration:} Average head movement speed (m/s) captured users' physical engagement and spatial exploration~\cite{YAREMYCH2019103845, Hu2023}. Higher values generally indicate active exploration, while movement variability may also reflect affective stance toward agents~\cite{McCall2017}.

\textbf{Verbal Interaction:} Binary verbal interaction (yes/no) with the AI tutor was used to assess social engagement. Prior work links interaction frequency and quality to user engagement and perceived social presence~\cite{FEINE2019138, Poivet2023}.

Additionally, we included exploratory post-assessment questions on participants’ attitudes, cultural interest, and subjective knowledge gains adapted from prior work~\cite{Konstantakis2020, Melo2023}. Given their supplementary role, these measures are detailed in the appendix.

\begin{table}[htb!]
\centering
\caption{Overview of multimodal metrics used in this study}
\resizebox{\columnwidth}{!}{%
\begin{tabular}{ll}
\toprule
\textbf{Modality Types} & \textbf{Dependent Variable} \\
\midrule
\textbf{Temporal Metric} & Total Experiment Time (seconds) \\
\textbf{Visual Attention} & Relative Avatar Dwell Time (\%) \\
\textbf{Physical Exploration} & Average Head Movement Speed (m/s) \\
\textbf{Verbal Interaction} & Interaction with AI (Yes/No) \\
\bottomrule
\end{tabular}%
}
\label{tab:metrics}
\end{table}

\subsection{Data Processing}

Collected data were processed using Python 3.9.6~\cite{pythonDocumentation}. Participants whose eye-tracking data quality fell below 80\% were excluded to ensure robustness~\cite{kasneci2024introeyetracking}. Eye-tracking data were analyzed by identifying gaze targets through a gaze-ray casting method~\cite{Bozkir2021, AlGhamdi2019}, mapping each gaze point to predefined AOIs (i.e., the virtual tutor (avatar), task-related surfaces, posters, chat interface, and the environment). Visual attention was quantified using dwell time, which is defined as the total duration spent gazing within each AOI during the session. Physical exploration was assessed through average head movement speed, calculated as the mean of instantaneous head velocities obtained by dividing the Euclidean distances between consecutive head positions (in meters) by the corresponding time intervals (in seconds). Temporal metric was measured by total session duration, derived from gaze data timestamps. Additionally, verbal interaction was recorded as a binary metric indicating whether participants interacted verbally with the virtual tutor at least once (Yes/No). These metrics were aggregated at the participant level for statistical analysis.

\subsection{Analysis}

Statistical analyses were conducted using Python 3.9.6~\cite{pythonDocumentation} with SciPy 1.13.1~\cite{SciPy} and Pingouin 0.5.5~\cite{Pingouin}. Normality was assessed using the Shapiro–Wilk test~\cite{Razali2011}, and homogeneity of variances was tested using Levene’s test~\cite{gastwirth2009impact}. For measures meeting the assumptions of normality but violating variance homogeneity, we conducted Welch’s ANOVA, followed by Games–Howell post-hoc tests for pairwise comparisons. Effect sizes for Welch’s ANOVAs were reported using partial eta-squared (\(\eta_p^{2}\)). When assumptions of normality were violated, we conducted non-parametric Kruskal–Wallis tests followed by Dunn's post-hoc tests with Holm–Bonferroni corrections. Effect sizes for Kruskal–Wallis tests were reported using epsilon-squared (\(\varepsilon^{2}\)). Categorical data (i.e., verbal interaction) were analysed using chi-square tests of independence, with effect sizes reported as Cramér’s \(V\). All statistical tests used an alpha level of \(p < .05\) as the threshold for significance.

\section{Results}
Each adaptivity condition (No, Moderate, High) had a sample size of \(n = 18\). Results are presented across four dimensions: temporal metric, visual attention, physical exploration, and verbal interaction.

\subsection{Temporal Metric}

\begin{figure}[ht!]
\centering
\includegraphics[width=0.8\columnwidth]{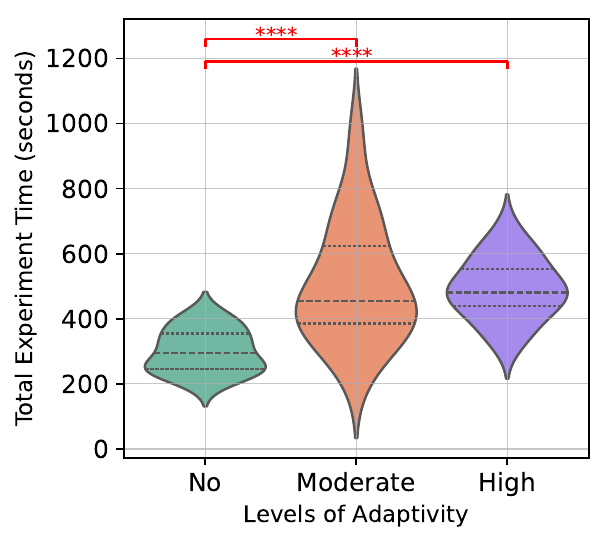}
\caption{Total experiment time (seconds) across adaptivity conditions (No, Moderate, High). Significance levels are indicated by ****, corresponding to \(p < .0001\).}
\Description{Total experiment time (seconds) across adaptivity conditions (No, Moderate, High). Significance levels are indicated by ****, corresponding to \(p < .0001\).}
\label{fig:totalExperimentTime}
\end{figure}

Figure~\ref{fig:totalExperimentTime} shows total experiment times across adaptivity conditions. Shapiro–Wilk tests indicated no significant deviation from normality (\(p > .05\)). However, Levene’s test revealed significant heterogeneity of variances (\(W(2, 51) = 4.27, p = .019\)), leading to the use of a Welch one-way ANOVA. Results showed a significant effect of adaptivity condition, \(F(2, 30.5) = 29.91, p < .001, \eta_p^{2} = .38\). Means (\(M\)), standard deviations (\(SD\)), and 95\% confidence intervals (\(CI\)) were: No Adaptivity (\(M = 300.00\), \(SD = 65.05\), 95\% \(CI\) [267.65, 332.35]), Moderate Adaptivity (\(M = 507.62\), \(SD = 182.97\), 95\% \(CI\) [416.64, 598.61]), and High Adaptivity (\(M = 492.24\), \(SD = 94.41\), 95\% \(CI\) [445.29, 539.19]). Games–Howell post-hoc tests indicated significant differences between No Adaptivity and both Moderate Adaptivity (mean difference = 207.62 s, 95\% \(CI\) [107.31, 307.93], \(p < .001\)) and High Adaptivity (mean difference = 192.24 s, 95\% \(CI\) [91.93, 292.55], \(p < .001\)). No significant difference was found between Moderate and High Adaptivity (\(p = .927\)).

\subsection{Visual Attention}

\begin{figure}[ht!]
\centering
\includegraphics[width=0.8\columnwidth]{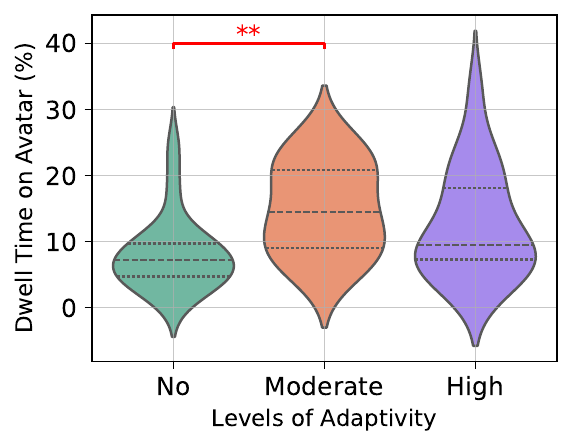}
\caption{Percentage of dwell time on the avatar across adaptivity conditions (No, Moderate, High). Significance levels are indicated by **, corresponding to \(p < .01\).}
\Description{Percentage of dwell time on the avatar across adaptivity conditions (No, Moderate, High). Significance levels are indicated by **, corresponding to \(p < .01\).}
\label{fig:avatarDwellTime}
\end{figure}

Figure~\ref{fig:avatarDwellTime} shows the percentage of session time spent dwelling on the avatar. Shapiro–Wilk tests revealed significant deviations from normality in No (\(W = 0.83, p = .004\)) and High (\(W = 0.89, p = .033\)) adaptivity conditions; thus, a Kruskal–Wallis test was used. Results showed a significant effect of adaptivity condition, \(H(2) = 10.11, p = .006, \varepsilon^{2} = .15\). Median percentages and interquartile ranges (IQR) were: No Adaptivity, 7.20\% (IQR = 4.71–9.68); Moderate Adaptivity, 14.47\% (IQR = 9.04–20.84); and High Adaptivity, 9.49\% (IQR = 7.32–18.13). Dunn’s post-hoc tests (Holm–Bonferroni correction) revealed a significant difference only between No and Moderate Adaptivity (\(p = .005\)); no significant differences occurred between No and High (\(p = .137\)) or Moderate and High Adaptivity (\(p = .178\)).

\subsection{Physical Exploration}

\begin{figure}[htp!]
\centering
\includegraphics[width=0.75\columnwidth]{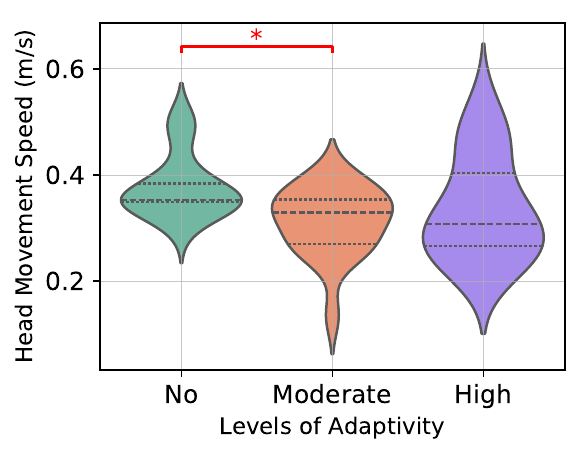}
\caption{Average head movement speed (m/s) across adaptivity conditions (No, Moderate, High). Significance levels are indicated by *, corresponding to \(p < .05\).}
\Description{Average head movement speed (m/s) across adaptivity conditions (No, Moderate, High). Significance levels are indicated by *, corresponding to \(p < .05\).}
\label{fig:headMovementSpeed}
\end{figure}

Figure~\ref{fig:headMovementSpeed} shows average head movement speed across adaptivity conditions. Shapiro–Wilk tests indicated a significant deviation from normality in the No Adaptivity condition (\(W = 0.80, p = .001\)); thus, a Kruskal–Wallis test was conducted. The analysis revealed a significant effect of adaptivity condition, \(H(2) = 7.74, p = .021, \varepsilon^{2} = .11\). Median speeds and IQR were: No Adaptivity, 0.353 m/s (IQR = 0.349–0.384); Moderate Adaptivity, 0.329 m/s (IQR = 0.270–0.354); and High Adaptivity, 0.307 m/s (IQR = 0.266–0.404). Dunn’s post-hoc tests (Holm–Bonferroni correction) indicated a significant difference between No and Moderate Adaptivity (\(p = .029\)); no significant differences emerged between No and High Adaptivity (\(p = .058\)) or Moderate and High Adaptivity (\(p = .687\)).

\subsection{Verbal Interaction}

\begin{figure}[htp!]
\centering
\includegraphics[width=0.8\columnwidth]{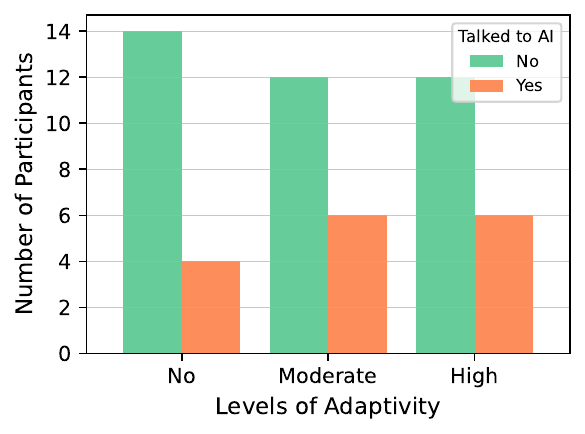}
\caption{Number of participants who verbally interacted with the AI (``Yes'') versus those who did not (``No'') across adaptivity conditions (No, Moderate, High). Differences across conditions were not statistically significant.}
\Description{Number of participants who verbally interacted with the AI (``Yes'') versus those who did not (``No'') across adaptivity conditions (No, Moderate, High). Differences across conditions were not statistically significant.}
\label{fig:talkedChatbot}
\end{figure}

Figure~\ref{fig:talkedChatbot} presents the proportions of participants who verbally interacted with the AI: No Adaptivity (22.2\%), Moderate Adaptivity (33.3\%), and High Adaptivity (33.3\%). A chi-square test revealed no significant association between adaptivity condition and verbal interaction, \(\chi^{2}(2, N = 54) = 0.71, p = .701, V = .08\). Adaptive guidance was primarily embedded in the AI's proactive instructions; consequently, verbal interaction was optional, and its absence does not imply a lack of adaptivity.

\section{Discussion}

To address our~\textbf{RQ}, our study investigated how adaptive generative AI affects learner engagement within a procedural VR scenario, specifically Neapolitan pizza-making. We found that moderate adaptivity, primarily tailored to real-time user choices, significantly enhanced visual attention and reduced extraneous exploratory behaviours compared to a non-adaptive baseline. Interestingly, introducing higher adaptivity, which additionally considered user demographics, did not yield further improvements, suggesting an optimal ``sweet spot'' for adaptivity intensity in procedural VR learning contexts.

We discuss our findings in three key areas: (1) the relationship between adaptivity levels and learner engagement, (2) the benefits of adopting a multimodal analytical framework compared to traditional HCI assessments, and (3) broader implications for designing adaptive AI tutors in procedural and ICH contexts. While our empirical case specifically examined Neapolitan pizza-making, the findings generalise to broader procedural ICH scenarios (e.g., traditional crafts and rituals) due to shared instructional characteristics and cultural depth. This generalisation highlights the potential of adaptive Gen-AI to effectively balance instructional guidance with learner autonomy across diverse cultural heritage applications.

\subsection{Adaptivity Levels and Learner Engagement}

Participants in both moderate and high adaptivity conditions had significantly longer session durations compared to the non-adaptive baseline. Although variability in AI response times may influence temporal metrics, session duration remains meaningful, as longer voluntary interactions generally signal sustained attention and motivation~\cite{Glas2015, Mayer_2014}. In the post-assessment, participants in the high-adaptivity condition reported significantly greater willingness to engage in cultural heritage activities after the intervention, such as visiting museums or volunteering at cultural events (Appendix Figure~\ref{fig:interest_violin}). However, because this result emerged from a single survey item rather than a broader validated scale, we interpret it cautiously as exploratory evidence supporting adaptivity’s potential to positively influence learner attitudes. This interpretation aligns with previous findings indicating that higher engagement with relevant content is associated with subsequent voluntary interactions~\cite{Epstein2022}. Combined with visual attention and exploratory behaviour metrics, our results suggest adaptive guidance is associated with higher learner engagement.

Interestingly, moderate adaptivity increased visual attention toward the avatar, whereas additional demographic-based adaptation (high adaptivity) provided no further improvement. This aligns with the `guidance-fading effect,' where excessive adaptation can reduce learner autonomy or intrinsic curiosity~\cite{Sweller2011}.

Reduced head movement in adaptive conditions further supports this interpretation, suggesting clearer instructional cues might have reduced the need for exploratory search. Prior VR studies similarly demonstrate that explicit instructional clarity can minimise cognitive distractions~\cite{DUBOVI2022104495} and reduce unnecessary exploratory head movements~\cite{YAREMYCH2019103845}.

Furthermore, we found no significant differences in verbal interactions. This aligns with prior research suggesting proactively embedded social and instructional cues in conversational agents can reduce users' reliance on explicit verbal communication~\cite{FEINE2019138}. Importantly, verbal interactions were optional in our study, as the tutor proactively provided guidance without requiring participants to initiate inquiries. Thus, moderate adaptivity emerges as optimal, effectively enhancing multimodal engagement without limiting learner agency.

Contrasting our findings with prior research on personalised content, which typically compares only binary conditions and their impact on attention and cognitive load~\cite{BANG2016867, Zander2015}, our study highlights a nuanced continuum of adaptivity. Specifically, our multimodal metrics reveal that moderate adaptivity achieves an optimal balance, effectively enhancing learner support without overwhelming learners, which is crucial in immersive procedural tasks. Thus, our findings extend previous research by demonstrating that the highest adaptation is not necessarily optimal in contexts requiring sustained cognitive engagement and learner autonomy.

\subsection{Benefits of Multimodal Evaluation}

Our multimodal evaluation framework, integrating eye-tracking, head movements, and verbal interactions, provides deeper insights than traditional single-modality or self-report approaches common in HCI research~\cite{Putze2020, Michael2020}. Traditional HCI metrics often struggle to capture dynamic user interactions, especially those involving adaptive AI systems, due to their reliance on isolated behavioural indicators or retrospective self-assessments~\cite{Xu2023, wangjiayin2024}.

By employing multimodal metrics, we identified nuanced patterns of cognitive engagement and social interaction. Eye-tracking highlighted sustained visual attention toward the avatar tutor, head movements signalled both exploratory behaviours and cognitive uncertainty, and optional verbal interactions captured explicit social engagement. This combination provides a holistic, unobtrusive, and real-time view of learning effectiveness in adaptive VR.

Broadly, this integrated multimodal approach contributes to HCI by demonstrating how behavioural analytics can effectively evaluate dynamic adaptive technologies. Our findings extend beyond procedural VR scenarios, providing practical guidance for designing more engaging and responsive user experiences across diverse interactive systems.

\subsection{Design Recommendations}

Based on our findings and discussion, we propose three key design recommendations:

\begin{enumerate}
    \item \textbf{Balance Adaptivity Level}\\
    Moderate adaptivity, which responds primarily to real-time user actions without demographic customisation, best supports learner engagement. We recommend defining an adaptivity threshold (e.g., limiting frequency or granularity of tailored hints) to guide learners effectively without diminishing their autonomy.

    \item \textbf{Leverage Multimodal Behavioural Cues}\\
    Eye-tracking and head-movement data reliably indicate confusion or loss of focus, allowing timely interventions. Systems should integrate real-time multimodal triggers (e.g., extended off-target gaze or rapid head movements) to provide targeted assistance precisely when learners require support.

    \item \textbf{Phase-Specific Adaptivity}\\
    Procedural learning alternates between structured steps, requiring explicit guidance, and open-ended exploration, benefiting from greater autonomy. We recommend dynamically varying adaptivity intensity: strong guidance during critical procedural phases (e.g., dough preparation) and lighter, more suggestive guidance during exploratory activities (e.g., ingredient selection).
\end{enumerate}

Our current prototype implements these recommendations in a simplified manner. Future designs can further incorporate inquiry-based learning principles to enhance learner agency.

\subsection{Limitations and Future Work}

Our findings have several limitations that inform future research directions. Firstly, we did not explicitly control participants' prior VR and cooking experiences, which could influence their engagement with the VR environment and interaction patterns with the adaptive AI tutor. Future studies should account for these background variables to isolate the effects of adaptivity more clearly. Additionally, the current prototype follows a linear interaction design, offering limited support for open-ended inquiry as emphasised by the CIL framework. Future iterations could incorporate more flexible task sequences and decision-making opportunities to better align with inquiry-driven approaches. Participants also reported insufficient interaction time with auxiliary cultural materials (i.e., posters) due to frequent interruptions from tutor dialogue. Subsequent designs might implement adaptive pacing mechanisms, such as dynamically adjusting tutor dialogue based on users' gaze fixation duration on these materials. Finally, inherent variability in generative AI responses could introduce inconsistencies in user experience and engagement metrics. Future research should systematically control this variability, possibly by standardising AI response timing or evaluating alternative response-generation methods.

\section{Conclusion}
This study investigated the effects of adaptive Gen-AI guidance on multimodal user engagement within a VR-based procedural learning context focused on intangible cultural heritage. Using the \textbf{Neapolitan Pizza VR} prototype in a controlled experiment, we demonstrated that moderate adaptivity driven by real-time learner actions optimally enhanced visual attention and reduced unnecessary exploratory behaviour. Our multimodal evaluation approach provided comprehensive insights into user engagement, addressing limitations of traditional HCI methods in evaluating dynamic adaptive systems. These findings underscore the importance of balancing adaptivity to effectively support learner engagement without compromising autonomy. Ultimately, this research offers practical guidelines for developing adaptive AI systems, with broad applicability to procedural learning scenarios in cultural heritage and beyond.

\section{Open Science}
To support research and development in VR and adaptive AI-based ICH education, we have made the source code and assets for Neapolitan Pizza VR available at the following link:~\url{https://anonymous.4open.science/r/NeapolitanPizzaVR}. 

\section{Safe and Responsible Innovation Statement}
This study received Institutional Review Board approval from [blind institute] and followed ethical research standards. Participants were informed about data handling procedures, and all multimodal data were anonymized, securely stored, and accessible only to authorized researchers. Recognizing ethical concerns related to adaptive AI, such as potential bias, lack of transparency, or misrepresentation of cultural content, we designed the VR experience to support equitable engagement and culturally respectful learning. We emphasize the importance of integrating adaptive AI responsibly by promoting fairness, protecting user privacy, and ensuring inclusivity in user interactions and educational outcomes.

\bibliographystyle{ACM-Reference-Format}
\bibliography{sample-base}


\begin{thebibliography}{80}

%%% the Web address.

\ifx \showCODEN    \undefined \def \showCODEN     #1{\unskip}     \fi
\ifx \showISBNx    \undefined \def \showISBNx     #1{\unskip}     \fi
\ifx \showISBNxiii \undefined \def \showISBNxiii  #1{\unskip}     \fi
\ifx \showISSN     \undefined \def \showISSN      #1{\unskip}     \fi
\ifx \showLCCN     \undefined \def \showLCCN      #1{\unskip}     \fi
\ifx \shownote     \undefined \def \shownote      #1{#1}          \fi
\ifx \showarticletitle \undefined \def \showarticletitle #1{#1}   \fi
\ifx \showURL      \undefined \def \showURL       {\relax}        \fi
% invisible to TeX
\providecommand\bibfield[2]{#2}
\providecommand\bibinfo[2]{#2}
\providecommand\natexlab[1]{#1}
\providecommand\showeprint[2][]{arXiv:#2}

\bibitem[Al~Ghamdi and Alhalabi(2019)]%
        {AlGhamdi2019}
\bibfield{author}{\bibinfo{person}{Najood Al~Ghamdi} {and} \bibinfo{person}{Wadee Alhalabi}.} \bibinfo{year}{2019}\natexlab{}.
\newblock \showarticletitle{Fixation Detection with Ray-casting in Immersive Virtual Reality}.
\newblock \bibinfo{journal}{\emph{International Journal of Advanced Computer Science and Applications}}  \bibinfo{volume}{10} (\bibinfo{date}{01} \bibinfo{year}{2019}).
\newblock
\href{https://doi.org/10.14569/IJACSA.2019.0100710}{doi:\nolinkurl{10.14569/IJACSA.2019.0100710}}


\bibitem[and(2006)]%
        {Ahmad15032006}
\bibfield{author}{\bibinfo{person}{Yahaya~Ahmad and}.} \bibinfo{year}{2006}\natexlab{}.
\newblock \showarticletitle{The Scope and Definitions of Heritage: From Tangible to Intangible}.
\newblock \bibinfo{journal}{\emph{International Journal of Heritage Studies}} \bibinfo{volume}{12}, \bibinfo{number}{3} (\bibinfo{year}{2006}), \bibinfo{pages}{292--300}.
\newblock
\href{https://doi.org/10.1080/13527250600604639}{doi:\nolinkurl{10.1080/13527250600604639}}


\bibitem[Aristidou et~al\mbox{.}(2021)]%
        {Aristidou2021}
\bibfield{author}{\bibinfo{person}{Andreas Aristidou}, \bibinfo{person}{Nefeli Andreou}, \bibinfo{person}{Loukas Charalambous}, \bibinfo{person}{Anastasios Yiannakidis}, {and} \bibinfo{person}{Yiorgos Chrysanthou}.} \bibinfo{year}{2021}\natexlab{}.
\newblock \showarticletitle{Virtual Dance Museum: the Case of Greek/Cypriot Folk Dancing}. In \bibinfo{booktitle}{\emph{Eurographics Workshop on Graphics and Cultural Heritage}}, \bibfield{editor}{\bibinfo{person}{Vedad Hulusic} {and} \bibinfo{person}{Alan Chalmers}} (Eds.). \bibinfo{publisher}{The Eurographics Association}, \bibinfo{address}{Eindhoven, Netherlands}, \bibinfo{pages}{1--10}.
\newblock
\showISBNx{978-3-03868-141-0}
\showISSN{2312-6124}
\href{https://doi.org/10.2312/gch.20211405}{doi:\nolinkurl{10.2312/gch.20211405}}


\bibitem[Bang and Wojdynski(2016)]%
        {BANG2016867}
\bibfield{author}{\bibinfo{person}{Hyejin Bang} {and} \bibinfo{person}{Bartosz~W. Wojdynski}.} \bibinfo{year}{2016}\natexlab{}.
\newblock \showarticletitle{Tracking users' visual attention and responses to personalized advertising based on task cognitive demand}.
\newblock \bibinfo{journal}{\emph{Computers in Human Behavior}}  \bibinfo{volume}{55} (\bibinfo{year}{2016}), \bibinfo{pages}{867--876}.
\newblock
\showISSN{0747-5632}
\href{https://doi.org/10.1016/j.chb.2015.10.025}{doi:\nolinkurl{10.1016/j.chb.2015.10.025}}


\bibitem[Beach and and(2019)]%
        {Beach20102019}
\bibfield{author}{\bibinfo{person}{Pamela Beach} {and} \bibinfo{person}{Jen~McConnel and}.} \bibinfo{year}{2019}\natexlab{}.
\newblock \showarticletitle{Eye tracking methodology for studying teacher learning: a review of the research}.
\newblock \bibinfo{journal}{\emph{International Journal of Research \& Method in Education}} \bibinfo{volume}{42}, \bibinfo{number}{5} (\bibinfo{year}{2019}), \bibinfo{pages}{485--501}.
\newblock
\href{https://doi.org/10.1080/1743727X.2018.1496415}{doi:\nolinkurl{10.1080/1743727X.2018.1496415}}
\showeprint{https://doi.org/10.1080/1743727X.2018.1496415}


\bibitem[Bekele et~al\mbox{.}(2018)]%
        {Bekele2018}
\bibfield{author}{\bibinfo{person}{Mafkereseb~Kassahun Bekele}, \bibinfo{person}{Roberto Pierdicca}, \bibinfo{person}{Emanuele Frontoni}, \bibinfo{person}{Eva~Savina Malinverni}, {and} \bibinfo{person}{James Gain}.} \bibinfo{year}{2018}\natexlab{}.
\newblock \showarticletitle{A Survey of Augmented, Virtual, and Mixed Reality for Cultural Heritage}.
\newblock \bibinfo{journal}{\emph{J. Comput. Cult. Herit.}} \bibinfo{volume}{11}, \bibinfo{number}{2}, Article \bibinfo{articleno}{7} (\bibinfo{date}{March} \bibinfo{year}{2018}), \bibinfo{numpages}{36}~pages.
\newblock
\showISSN{1556-4673}
\href{https://doi.org/10.1145/3145534}{doi:\nolinkurl{10.1145/3145534}}


\bibitem[Bernacki and Walkington(2018)]%
        {Bernacki2018}
\bibfield{author}{\bibinfo{person}{Matthew Bernacki} {and} \bibinfo{person}{Candace Walkington}.} \bibinfo{year}{2018}\natexlab{}.
\newblock \showarticletitle{The Role of Situational Interest in Personalized Learning}.
\newblock \bibinfo{journal}{\emph{Journal of Educational Psychology}}  \bibinfo{volume}{110} (\bibinfo{date}{03} \bibinfo{year}{2018}).
\newblock
\href{https://doi.org/10.1037/edu0000250}{doi:\nolinkurl{10.1037/edu0000250}}


\bibitem[Bieniek et~al\mbox{.}(2024)]%
        {Bieniek2024}
\bibfield{author}{\bibinfo{person}{J. Bieniek}, \bibinfo{person}{M. Rahouti}, {and} \bibinfo{person}{D.~C. Verma}.} \bibinfo{year}{2024}\natexlab{}.
\newblock \bibinfo{title}{Generative AI in Multimodal User Interfaces: Trends, Challenges, and Cross-Platform Adaptability}.
\newblock
\showeprint[arxiv]{2411.10234}~[cs.HC]
\urldef\tempurl%
\url{https://arxiv.org/abs/2411.10234}
\showURL{%
\tempurl}


\bibitem[Bozkir et~al\mbox{.}(2021)]%
        {Bozkir2021}
\bibfield{author}{\bibinfo{person}{Efe Bozkir}, \bibinfo{person}{Philipp Stark}, \bibinfo{person}{Hong Gao}, \bibinfo{person}{Lisa Hasenbein}, \bibinfo{person}{Jens-Uwe Hahn}, \bibinfo{person}{Enkelejda Kasneci}, {and} \bibinfo{person}{Richard Göllner}.} \bibinfo{year}{2021}\natexlab{}.
\newblock \showarticletitle{Exploiting Object-of-Interest Information to Understand Attention in VR Classrooms}. In \bibinfo{booktitle}{\emph{2021 IEEE Virtual Reality and 3D User Interfaces (VR)}}. \bibinfo{publisher}{IEEE}, \bibinfo{address}{Lisboa, Portugal}, \bibinfo{pages}{597--605}.
\newblock
\href{https://doi.org/10.1109/VR50410.2021.00085}{doi:\nolinkurl{10.1109/VR50410.2021.00085}}


\bibitem[Brady(2011)]%
        {Brady2011}
\bibfield{author}{\bibinfo{person}{Jennifer Brady}.} \bibinfo{year}{2011}\natexlab{}.
\newblock \showarticletitle{Cooking as Inquiry: A Method to Stir Up Prevailing Ways of Knowing Food, Body, and Identity}.
\newblock \bibinfo{journal}{\emph{International Journal of Qualitative Methods}} \bibinfo{volume}{10}, \bibinfo{number}{4} (\bibinfo{year}{2011}), \bibinfo{pages}{321--334}.
\newblock
\href{https://doi.org/10.1177/160940691101000402}{doi:\nolinkurl{10.1177/160940691101000402}}


\bibitem[Brako(2024)]%
        {Brako2024}
\bibfield{author}{\bibinfo{person}{Morris Brako}.} \bibinfo{year}{2024}\natexlab{}.
\newblock \showarticletitle{Preserving culinary heritage: Challenges faced by Takoradi technical university food service lab students}.
\newblock \bibinfo{journal}{\emph{World Journal of Advanced Research and Reviews}}  \bibinfo{volume}{21} (\bibinfo{date}{03} \bibinfo{year}{2024}), \bibinfo{pages}{1536--1545}.
\newblock
\href{https://doi.org/10.30574/wjarr.2024.21.3.0882}{doi:\nolinkurl{10.30574/wjarr.2024.21.3.0882}}


\bibitem[Brulotte and Di~Giovine(2016)]%
        {brulotte2016edible}
\bibfield{author}{\bibinfo{person}{Ronda~L Brulotte} {and} \bibinfo{person}{Michael~A Di~Giovine}.} \bibinfo{year}{2016}\natexlab{}.
\newblock \bibinfo{booktitle}{\emph{Edible identities: Food as cultural heritage}}.
\newblock \bibinfo{publisher}{Routledge}, \bibinfo{address}{London, UK}.
\newblock


\bibitem[Carrozzino et~al\mbox{.}(2011)]%
        {CARROZZINO201182}
\bibfield{author}{\bibinfo{person}{Marcello Carrozzino}, \bibinfo{person}{Alessandra Scucces}, \bibinfo{person}{Rosario Leonardi}, \bibinfo{person}{Chiara Evangelista}, {and} \bibinfo{person}{Massimo Bergamasco}.} \bibinfo{year}{2011}\natexlab{}.
\newblock \showarticletitle{Virtually preserving the intangible heritage of artistic handicraft}.
\newblock \bibinfo{journal}{\emph{Journal of Cultural Heritage}} \bibinfo{volume}{12}, \bibinfo{number}{1} (\bibinfo{year}{2011}), \bibinfo{pages}{82--87}.
\newblock
\showISSN{1296-2074}
\href{https://doi.org/10.1016/j.culher.2010.10.002}{doi:\nolinkurl{10.1016/j.culher.2010.10.002}}


\bibitem[Ceccarini(2011)]%
        {ceccarini2011pizza}
\bibfield{author}{\bibinfo{person}{Rossella Ceccarini}.} \bibinfo{year}{2011}\natexlab{}.
\newblock \bibinfo{booktitle}{\emph{Pizza and pizza chefs in Japan: A case of culinary globalization}}. Vol.~\bibinfo{volume}{31}.
\newblock \bibinfo{publisher}{Brill Academic Publishers}, \bibinfo{address}{Leiden; Boston}.
\newblock


\bibitem[Dagnino et~al\mbox{.}(2015)]%
        {Dagnino2015}
\bibfield{author}{\bibinfo{person}{Francesca~Maria Dagnino}, \bibinfo{person}{Francesca Pozzi}, \bibinfo{person}{Erdal Yilmaz}, \bibinfo{person}{Nikos Grammalidis}, \bibinfo{person}{Kosmas Dimitropoulos}, {and} \bibinfo{person}{Filareti Tsalakanidou}.} \bibinfo{year}{2015}\natexlab{}.
\newblock \showarticletitle{Designing Serious Games for ICH education}. In \bibinfo{booktitle}{\emph{2015 Digital Heritage}}, Vol.~\bibinfo{volume}{2}. \bibinfo{publisher}{IEEE}, \bibinfo{address}{Granada, Spain}, \bibinfo{pages}{615--618}.
\newblock
\href{https://doi.org/10.1109/DigitalHeritage.2015.7419581}{doi:\nolinkurl{10.1109/DigitalHeritage.2015.7419581}}


\bibitem[Damala et~al\mbox{.}(2012)]%
        {Damala2012}
\bibfield{author}{\bibinfo{person}{Areti Damala}, \bibinfo{person}{Nenad Stojanovic}, \bibinfo{person}{Tobias Schuchert}, \bibinfo{person}{Jorge Moragues}, \bibinfo{person}{Ana Cabrera}, {and} \bibinfo{person}{Kiel Gilleade}.} \bibinfo{year}{2012}\natexlab{}.
\newblock \showarticletitle{Adaptive Augmented Reality for Cultural Heritage: ARtSENSE Project}. In \bibinfo{booktitle}{\emph{Progress in Cultural Heritage Preservation}}, \bibfield{editor}{\bibinfo{person}{Marinos Ioannides}, \bibinfo{person}{Dieter Fritsch}, \bibinfo{person}{Johanna Leissner}, \bibinfo{person}{Rob Davies}, \bibinfo{person}{Fabio Remondino}, {and} \bibinfo{person}{Rossella Caffo}} (Eds.). \bibinfo{publisher}{Springer Berlin Heidelberg}, \bibinfo{address}{Berlin, Heidelberg}, \bibinfo{pages}{746--755}.
\newblock
\showISBNx{978-3-642-34234-9}


\bibitem[Di~Fiore(2019)]%
        {di2020heritage}
\bibfield{author}{\bibinfo{person}{Laura Di~Fiore}.} \bibinfo{year}{2019}\natexlab{}.
\newblock \showarticletitle{Heritage and food history: A critical assessment}.
\newblock In \bibinfo{booktitle}{\emph{Food heritage and nationalism in Europe}}. \bibinfo{publisher}{Routledge}, \bibinfo{address}{Abingdon}.
\newblock


\bibitem[Do et~al\mbox{.}(2025)]%
        {Do2024}
\bibfield{author}{\bibinfo{person}{Tiffany~D. Do}, \bibinfo{person}{Usama~Bin Shafqat}, \bibinfo{person}{Elsie Ling}, {and} \bibinfo{person}{Nikhil Sarda}.} \bibinfo{year}{2025}\natexlab{}.
\newblock \showarticletitle{PAIGE: Examining Learning Outcomes and Experiences with Personalized AI-Generated Educational Podcasts}. In \bibinfo{booktitle}{\emph{Proceedings of the 2025 CHI Conference on Human Factors in Computing Systems}} \emph{(\bibinfo{series}{CHI '25})}. \bibinfo{publisher}{Association for Computing Machinery}, \bibinfo{address}{New York, NY, USA}, Article \bibinfo{articleno}{896}, \bibinfo{numpages}{12}~pages.
\newblock
\showISBNx{9798400713941}
\href{https://doi.org/10.1145/3706598.3713460}{doi:\nolinkurl{10.1145/3706598.3713460}}


\bibitem[Dong et~al\mbox{.}(2021)]%
        {Dong2021}
\bibfield{author}{\bibinfo{person}{Beibei Dong}, \bibinfo{person}{Shangshi Pan}, {and} \bibinfo{person}{RongRong Fu}.} \bibinfo{year}{2021}\natexlab{}.
\newblock \showarticletitle{Design and Research on the Virtual Simulation Teaching Platform of Shanghai Jade Carving Techniques Based on Unity 3D Technology}. In \bibinfo{booktitle}{\emph{Virtual, Augmented and Mixed Reality}}, \bibfield{editor}{\bibinfo{person}{Jessie Y.~C. Chen} {and} \bibinfo{person}{Gino Fragomeni}} (Eds.). \bibinfo{publisher}{Springer International Publishing}, \bibinfo{address}{Cham}, \bibinfo{pages}{571--581}.
\newblock
\showISBNx{978-3-030-77599-5}


\bibitem[Dubovi(2022)]%
        {DUBOVI2022104495}
\bibfield{author}{\bibinfo{person}{Ilana Dubovi}.} \bibinfo{year}{2022}\natexlab{}.
\newblock \showarticletitle{Cognitive and emotional engagement while learning with VR: The perspective of multimodal methodology}.
\newblock \bibinfo{journal}{\emph{Computers \& Education}}  \bibinfo{volume}{183} (\bibinfo{year}{2022}), \bibinfo{pages}{104495}.
\newblock
\showISSN{0360-1315}
\href{https://doi.org/10.1016/j.compedu.2022.104495}{doi:\nolinkurl{10.1016/j.compedu.2022.104495}}


\bibitem[Epstein et~al\mbox{.}(2022)]%
        {Epstein2022}
\bibfield{author}{\bibinfo{person}{Ziv Epstein}, \bibinfo{person}{Hause Lin}, {and} \bibinfo{person}{Gordon Pennycook}.} \bibinfo{year}{2022}\natexlab{}.
\newblock \bibinfo{title}{Quantifying attention via dwell time and engagement in a social media browsing environment}.
\newblock
\href{https://doi.org/10.48550/arXiv.2209.10464}{doi:\nolinkurl{10.48550/arXiv.2209.10464}}


\bibitem[Er-Rafyg et~al\mbox{.}(2024)]%
        {ErRafyg2024}
\bibfield{author}{\bibinfo{person}{Aicha Er-Rafyg}, \bibinfo{person}{Hajar Zankadi}, {and} \bibinfo{person}{Abdellah Idrissi}.} \bibinfo{year}{2024}\natexlab{}.
\newblock \bibinfo{booktitle}{\emph{AI in Adaptive Learning: Challenges and Opportunities}}.
\newblock \bibinfo{publisher}{Springer Nature Switzerland}, \bibinfo{address}{Cham}, \bibinfo{pages}{329--342}.
\newblock
\showISBNx{978-3-031-65038-3}
\href{https://doi.org/10.1007/978-3-031-65038-3_26}{doi:\nolinkurl{10.1007/978-3-031-65038-3_26}}


\bibitem[Feine et~al\mbox{.}(2019)]%
        {FEINE2019138}
\bibfield{author}{\bibinfo{person}{Jasper Feine}, \bibinfo{person}{Ulrich Gnewuch}, \bibinfo{person}{Stefan Morana}, {and} \bibinfo{person}{Alexander Maedche}.} \bibinfo{year}{2019}\natexlab{}.
\newblock \showarticletitle{A Taxonomy of Social Cues for Conversational Agents}.
\newblock \bibinfo{journal}{\emph{International Journal of Human-Computer Studies}}  \bibinfo{volume}{132} (\bibinfo{year}{2019}), \bibinfo{pages}{138--161}.
\newblock
\showISSN{1071-5819}
\href{https://doi.org/10.1016/j.ijhcs.2019.07.009}{doi:\nolinkurl{10.1016/j.ijhcs.2019.07.009}}


\bibitem[Garbutt et~al\mbox{.}(2020)]%
        {Michael2020}
\bibfield{author}{\bibinfo{person}{Michael Garbutt}, \bibinfo{person}{Scott East}, \bibinfo{person}{Branka Spehar}, \bibinfo{person}{Vicente Estrada-Gonzalez}, \bibinfo{person}{Brooke Carson-Ewart}, {and} \bibinfo{person}{Josephine Touma}.} \bibinfo{year}{2020}\natexlab{}.
\newblock \showarticletitle{The Embodied Gaze: Exploring Applications for Mobile Eye Tracking in the Art Museum}.
\newblock \bibinfo{journal}{\emph{Visitor Studies}} \bibinfo{volume}{23}, \bibinfo{number}{1} (\bibinfo{year}{2020}), \bibinfo{pages}{82--100}.
\newblock
\href{https://doi.org/10.1080/10645578.2020.1750271}{doi:\nolinkurl{10.1080/10645578.2020.1750271}}


\bibitem[Gastwirth et~al\mbox{.}(2009)]%
        {gastwirth2009impact}
\bibfield{author}{\bibinfo{person}{Joseph~L Gastwirth}, \bibinfo{person}{Yulia~R Gel}, {and} \bibinfo{person}{Weiwen Miao}.} \bibinfo{year}{2009}\natexlab{}.
\newblock \showarticletitle{The impact of Levene’s test of equality of variances on statistical theory and practice}.
\newblock \bibinfo{journal}{\emph{Statist. Sci.}} \bibinfo{volume}{24}, \bibinfo{number}{3} (\bibinfo{year}{2009}), \bibinfo{pages}{343--360}.
\newblock


\bibitem[Glas and Pelachaud(2015)]%
        {Glas2015}
\bibfield{author}{\bibinfo{person}{Nadine Glas} {and} \bibinfo{person}{Catherine Pelachaud}.} \bibinfo{year}{2015}\natexlab{}.
\newblock \showarticletitle{Definitions of engagement in human-agent interaction}. In \bibinfo{booktitle}{\emph{2015 International Conference on Affective Computing and Intelligent Interaction (ACII)}}. \bibinfo{publisher}{IEEE}, \bibinfo{address}{Xi'an, China}, \bibinfo{pages}{944--949}.
\newblock
\href{https://doi.org/10.1109/ACII.2015.7344688}{doi:\nolinkurl{10.1109/ACII.2015.7344688}}


\bibitem[Guo et~al\mbox{.}(2024)]%
        {guo2024enhancing}
\bibfield{author}{\bibinfo{person}{Hua Guo}, \bibinfo{person}{Weiqian Yi}, {and} \bibinfo{person}{Kecheng Liu}.} \bibinfo{year}{2024}\natexlab{}.
\newblock \showarticletitle{Enhancing Constructivist Learning: The Role of Generative AI in Personalised Learning Experiences}. In \bibinfo{booktitle}{\emph{Proceedings of the 26th International Conference on Enterprise Information Systems (ICEIS 2024) - Volume 1}}. \bibinfo{publisher}{SCITEPRESS – Science and Technology Publications, Lda.}, \bibinfo{address}{Benidorm, Spain}, \bibinfo{pages}{767--770}.
\newblock
\href{https://doi.org/10.5220/0012688700003690}{doi:\nolinkurl{10.5220/0012688700003690}}


\bibitem[Hidi and and(2006)]%
        {Hidi01062006}
\bibfield{author}{\bibinfo{person}{Suzanne Hidi} {and} \bibinfo{person}{K.~Ann~Renninger and}.} \bibinfo{year}{2006}\natexlab{}.
\newblock \showarticletitle{The Four-Phase Model of Interest Development}.
\newblock \bibinfo{journal}{\emph{Educational Psychologist}} \bibinfo{volume}{41}, \bibinfo{number}{2} (\bibinfo{year}{2006}), \bibinfo{pages}{111--127}.
\newblock
\href{https://doi.org/10.1207/s15326985ep4102\_4}{doi:\nolinkurl{10.1207/s15326985ep4102\_4}}


\bibitem[Hou et~al\mbox{.}(2022)]%
        {Hou2022}
\bibfield{author}{\bibinfo{person}{Yumeng Hou}, \bibinfo{person}{Sarah Kenderdine}, \bibinfo{person}{Davide Picca}, \bibinfo{person}{Mattia Egloff}, {and} \bibinfo{person}{Alessandro Adamou}.} \bibinfo{year}{2022}\natexlab{}.
\newblock \showarticletitle{Digitizing Intangible Cultural Heritage Embodied: State of the Art}.
\newblock \bibinfo{journal}{\emph{J. Comput. Cult. Herit.}} \bibinfo{volume}{15}, \bibinfo{number}{3}, Article \bibinfo{articleno}{55} (\bibinfo{date}{Sept.} \bibinfo{year}{2022}), \bibinfo{numpages}{20}~pages.
\newblock
\showISSN{1556-4673}
\href{https://doi.org/10.1145/3494837}{doi:\nolinkurl{10.1145/3494837}}


\bibitem[Hu et~al\mbox{.}(2023)]%
        {Hu2023}
\bibfield{author}{\bibinfo{person}{Zhiming Hu}, \bibinfo{person}{Andreas Bulling}, \bibinfo{person}{Sheng Li}, {and} \bibinfo{person}{Guoping Wang}.} \bibinfo{year}{2023}\natexlab{}.
\newblock \showarticletitle{EHTask: Recognizing User Tasks From Eye and Head Movements in Immersive Virtual Reality}.
\newblock \bibinfo{journal}{\emph{IEEE Transactions on Visualization and Computer Graphics}} \bibinfo{volume}{29}, \bibinfo{number}{4} (\bibinfo{year}{2023}), \bibinfo{pages}{1992--2004}.
\newblock
\href{https://doi.org/10.1109/TVCG.2021.3138902}{doi:\nolinkurl{10.1109/TVCG.2021.3138902}}


\bibitem[Huang et~al\mbox{.}(2023)]%
        {huang2023surveyhallucinationlargelanguage}
\bibfield{author}{\bibinfo{person}{Lei Huang}, \bibinfo{person}{Weijiang Yu}, \bibinfo{person}{Weitao Ma}, \bibinfo{person}{Weihong Zhong}, \bibinfo{person}{Zhangyin Feng}, \bibinfo{person}{Haotian Wang}, \bibinfo{person}{Qianglong Chen}, \bibinfo{person}{Weihua Peng}, \bibinfo{person}{Xiaocheng Feng}, \bibinfo{person}{Bing Qin}, {and} \bibinfo{person}{Ting Liu}.} \bibinfo{year}{2023}\natexlab{}.
\newblock \bibinfo{title}{A Survey on Hallucination in Large Language Models: Principles, Taxonomy, Challenges, and Open Questions}.
\newblock
\showeprint[arxiv]{2311.05232}~[cs.CL]
\urldef\tempurl%
\url{https://arxiv.org/abs/2311.05232}
\showURL{%
\tempurl}


\bibitem[Huang et~al\mbox{.}(2012)]%
        {Huang2012}
\bibfield{author}{\bibinfo{person}{Yong-Ming Huang}, \bibinfo{person}{C.-H Liu}, {and} \bibinfo{person}{C.-Y Lee}.} \bibinfo{year}{2012}\natexlab{}.
\newblock \showarticletitle{Designing a personalized guide recommendation system to mitigate information overload in museum learning}.
\newblock \bibinfo{journal}{\emph{Educational Technology and Society}}  \bibinfo{volume}{15} (\bibinfo{date}{01} \bibinfo{year}{2012}), \bibinfo{pages}{150--166}.
\newblock


\bibitem[Kabudi et~al\mbox{.}(2021)]%
        {KABUDI2021100017}
\bibfield{author}{\bibinfo{person}{Tumaini Kabudi}, \bibinfo{person}{Ilias Pappas}, {and} \bibinfo{person}{Dag~Håkon Olsen}.} \bibinfo{year}{2021}\natexlab{}.
\newblock \showarticletitle{AI-enabled adaptive learning systems: A systematic mapping of the literature}.
\newblock \bibinfo{journal}{\emph{Computers and Education: Artificial Intelligence}}  \bibinfo{volume}{2} (\bibinfo{year}{2021}), \bibinfo{pages}{100017}.
\newblock
\showISSN{2666-920X}
\href{https://doi.org/10.1016/j.caeai.2021.100017}{doi:\nolinkurl{10.1016/j.caeai.2021.100017}}


\bibitem[Kasneci et~al\mbox{.}(2024)]%
        {kasneci2024introeyetracking}
\bibfield{author}{\bibinfo{person}{Enkelejda Kasneci}, \bibinfo{person}{Hong Gao}, \bibinfo{person}{Suleyman Ozdel}, \bibinfo{person}{Virmarie Maquiling}, \bibinfo{person}{Enkeleda Thaqi}, \bibinfo{person}{Carrie Lau}, \bibinfo{person}{Yao Rong}, \bibinfo{person}{Gjergji Kasneci}, {and} \bibinfo{person}{Efe Bozkir}.} \bibinfo{year}{2024}\natexlab{}.
\newblock \bibinfo{title}{Introduction to Eye Tracking: A Hands-On Tutorial for Students and Practitioners}.
\newblock
\showeprint[arxiv]{2404.15435}~[cs.HC]
\urldef\tempurl%
\url{https://arxiv.org/abs/2404.15435}
\showURL{%
\tempurl}


\bibitem[Kasneci et~al\mbox{.}(2023)]%
        {KASNECI2023102274}
\bibfield{author}{\bibinfo{person}{Enkelejda Kasneci}, \bibinfo{person}{Kathrin Sessler}, \bibinfo{person}{Stefan Küchemann}, \bibinfo{person}{Maria Bannert}, \bibinfo{person}{Daryna Dementieva}, \bibinfo{person}{Frank Fischer}, \bibinfo{person}{Urs Gasser}, \bibinfo{person}{Georg Groh}, \bibinfo{person}{Stephan Günnemann}, \bibinfo{person}{Eyke Hüllermeier}, \bibinfo{person}{Stephan Krusche}, \bibinfo{person}{Gitta Kutyniok}, \bibinfo{person}{Tilman Michaeli}, \bibinfo{person}{Claudia Nerdel}, \bibinfo{person}{Jürgen Pfeffer}, \bibinfo{person}{Oleksandra Poquet}, \bibinfo{person}{Michael Sailer}, \bibinfo{person}{Albrecht Schmidt}, \bibinfo{person}{Tina Seidel}, \bibinfo{person}{Matthias Stadler}, \bibinfo{person}{Jochen Weller}, \bibinfo{person}{Jochen Kuhn}, {and} \bibinfo{person}{Gjergji Kasneci}.} \bibinfo{year}{2023}\natexlab{}.
\newblock \showarticletitle{ChatGPT for good? On opportunities and challenges of large language models for education}.
\newblock \bibinfo{journal}{\emph{Learning and Individual Differences}}  \bibinfo{volume}{103} (\bibinfo{year}{2023}), \bibinfo{pages}{102274}.
\newblock
\showISSN{1041-6080}
\href{https://doi.org/10.1016/j.lindif.2023.102274}{doi:\nolinkurl{10.1016/j.lindif.2023.102274}}


\bibitem[Kenderdine et~al\mbox{.}(2014)]%
        {Kenderdine2014}
\bibfield{author}{\bibinfo{person}{Sarah Kenderdine}, \bibinfo{person}{Leith K.~Y. Chan}, {and} \bibinfo{person}{Jeffrey Shaw}.} \bibinfo{year}{2014}\natexlab{}.
\newblock \showarticletitle{Pure Land: Futures for Embodied Museography}.
\newblock \bibinfo{journal}{\emph{J. Comput. Cult. Herit.}} \bibinfo{volume}{7}, \bibinfo{number}{2}, Article \bibinfo{articleno}{8} (\bibinfo{date}{jun} \bibinfo{year}{2014}), \bibinfo{numpages}{15}~pages.
\newblock
\showISSN{1556-4673}
\href{https://doi.org/10.1145/2614567}{doi:\nolinkurl{10.1145/2614567}}


\bibitem[Kim et~al\mbox{.}(2021)]%
        {Kim2021}
\bibfield{author}{\bibinfo{person}{Kyoung~Jin Kim}, \bibinfo{person}{Jiyoon Yoon}, {and} \bibinfo{person}{Min-Kyung Han}.} \bibinfo{year}{2021}\natexlab{}.
\newblock \showarticletitle{Young chefs in the classroom: promoting scientific process skills and healthy eating habits through an inquiry-based cooking project}.
\newblock \bibinfo{journal}{\emph{International Journal of Early Years Education}}  \bibinfo{volume}{30} (\bibinfo{date}{08} \bibinfo{year}{2021}), \bibinfo{pages}{1--10}.
\newblock
\href{https://doi.org/10.1080/09669760.2021.1892597}{doi:\nolinkurl{10.1080/09669760.2021.1892597}}


\bibitem[Konstantakis and Caridakis(2020)]%
        {Konstantakis2020}
\bibfield{author}{\bibinfo{person}{Markos Konstantakis} {and} \bibinfo{person}{George Caridakis}.} \bibinfo{year}{2020}\natexlab{}.
\newblock \showarticletitle{Adding Culture to UX: UX Research Methodologies and Applications in Cultural Heritage}.
\newblock \bibinfo{journal}{\emph{Journal on Computing and Cultural Heritage}}  \bibinfo{volume}{13} (\bibinfo{date}{02} \bibinfo{year}{2020}), \bibinfo{pages}{1--17}.
\newblock
\href{https://doi.org/10.1145/3354002}{doi:\nolinkurl{10.1145/3354002}}


\bibitem[Konstantakis et~al\mbox{.}(2022)]%
        {digital2030020}
\bibfield{author}{\bibinfo{person}{Markos Konstantakis}, \bibinfo{person}{Yannis Christodoulou}, \bibinfo{person}{Georgios Alexandridis}, \bibinfo{person}{Alexandros Teneketzis}, {and} \bibinfo{person}{George Caridakis}.} \bibinfo{year}{2022}\natexlab{}.
\newblock \showarticletitle{ACUX Typology: A Harmonisation of Cultural-Visitor Typologies for Multi-Profile Classification}.
\newblock \bibinfo{journal}{\emph{Digital}} \bibinfo{volume}{2}, \bibinfo{number}{3} (\bibinfo{year}{2022}), \bibinfo{pages}{365--378}.
\newblock
\showISSN{2673-6470}
\href{https://doi.org/10.3390/digital2030020}{doi:\nolinkurl{10.3390/digital2030020}}


\bibitem[Lau et~al\mbox{.}(2024)]%
        {lau2024evaluating}
\bibfield{author}{\bibinfo{person}{Ka~Hei~Carrie Lau}, \bibinfo{person}{Efe Bozkir}, \bibinfo{person}{Hong Gao}, {and} \bibinfo{person}{Enkelejda Kasneci}.} \bibinfo{year}{2024}\natexlab{}.
\newblock \bibinfo{title}{Evaluating Usability and Engagement of Large Language Models in Virtual Reality for Traditional Scottish Curling}.
\newblock
\showeprint[arxiv]{2408.09285}
\urldef\tempurl%
\url{https://arxiv.org/abs/2408.09285}
\showURL{%
\tempurl}


\bibitem[Lee(2023)]%
        {LEE20231}
\bibfield{author}{\bibinfo{person}{Kai-Sean Lee}.} \bibinfo{year}{2023}\natexlab{}.
\newblock \showarticletitle{Cooking up food memories: A taste of intangible cultural heritage}.
\newblock \bibinfo{journal}{\emph{Journal of Hospitality and Tourism Management}}  \bibinfo{volume}{54} (\bibinfo{year}{2023}), \bibinfo{pages}{1--9}.
\newblock
\showISSN{1447-6770}
\href{https://doi.org/10.1016/j.jhtm.2022.11.005}{doi:\nolinkurl{10.1016/j.jhtm.2022.11.005}}


\bibitem[Leong et~al\mbox{.}(2024)]%
        {Leong2024}
\bibfield{author}{\bibinfo{person}{Joanne Leong}, \bibinfo{person}{Pat Pataranutaporn}, \bibinfo{person}{Valdemar Danry}, \bibinfo{person}{Florian Perteneder}, \bibinfo{person}{Yaoli Mao}, {and} \bibinfo{person}{Pattie Maes}.} \bibinfo{year}{2024}\natexlab{}.
\newblock \showarticletitle{Putting Things into Context: Generative AI-Enabled Context Personalization for Vocabulary Learning Improves Learning Motivation}. In \bibinfo{booktitle}{\emph{Proceedings of the CHI Conference on Human Factors in Computing Systems}} (Honolulu, HI, USA) \emph{(\bibinfo{series}{CHI '24})}. \bibinfo{publisher}{Association for Computing Machinery}, \bibinfo{address}{New York, NY, USA}, Article \bibinfo{articleno}{677}, \bibinfo{numpages}{15}~pages.
\newblock
\showISBNx{9798400703300}
\href{https://doi.org/10.1145/3613904.3642393}{doi:\nolinkurl{10.1145/3613904.3642393}}


\bibitem[L{\'e}vi‐Strauss(2012)]%
        {LviStrauss2012TheCT}
\bibfield{author}{\bibinfo{person}{Claude L{\'e}vi‐Strauss}.} \bibinfo{year}{2012}\natexlab{}.
\newblock \showarticletitle{The Culinary Triangle}.
\newblock In \bibinfo{booktitle}{\emph{Food and Culture: A Reader}}, \bibfield{editor}{\bibinfo{person}{Carole Counihan} {and} \bibinfo{person}{Penny Van~Esterik}} (Eds.). \bibinfo{publisher}{Routledge}, \bibinfo{address}{New York, NY, USA}, \bibinfo{pages}{40--47}.
\newblock
\href{https://doi.org/10.4324/9781315680347-3}{doi:\nolinkurl{10.4324/9781315680347-3}}


\bibitem[Li et~al\mbox{.}(2022)]%
        {LiNa2022}
\bibfield{author}{\bibinfo{person}{Na Li}, \bibinfo{person}{Shanshan Zhang}, \bibinfo{person}{Lei Xia}, {and} \bibinfo{person}{Yue Wu}.} \bibinfo{year}{2022}\natexlab{}.
\newblock \showarticletitle{Investigating the Visual Behavior Characteristics of Architectural Heritage Using Eye-Tracking}.
\newblock \bibinfo{journal}{\emph{Buildings}}  \bibinfo{volume}{12} (\bibinfo{date}{07} \bibinfo{year}{2022}), \bibinfo{pages}{1058}.
\newblock
\href{https://doi.org/10.3390/buildings12071058}{doi:\nolinkurl{10.3390/buildings12071058}}


\bibitem[Loomis et~al\mbox{.}(2008)]%
        {Loomis2008}
\bibfield{author}{\bibinfo{person}{Jack~M Loomis}, \bibinfo{person}{Jonathan~W Kelly}, \bibinfo{person}{Matthias Pusch}, \bibinfo{person}{Jeremy~N Bailenson}, {and} \bibinfo{person}{Andrew~C Beall}.} \bibinfo{year}{2008}\natexlab{}.
\newblock \showarticletitle{Psychophysics of Perceiving Eye-Gaze and Head Direction with Peripheral Vision: Implications for the Dynamics of Eye-Gaze Behavior}.
\newblock \bibinfo{journal}{\emph{Perception}} \bibinfo{volume}{37}, \bibinfo{number}{9} (\bibinfo{year}{2008}), \bibinfo{pages}{1443--1457}.
\newblock
\href{https://doi.org/10.1068/p5896}{doi:\nolinkurl{10.1068/p5896}}
\newblock
\shownote{PMID: 18986070}.


\bibitem[Mayer(2014)]%
        {Mayer_2014}
\bibfield{author}{\bibinfo{person}{Richard~E. Mayer}.} \bibinfo{year}{2014}\natexlab{}.
\newblock \bibinfo{booktitle}{\emph{Cognitive Theory of Multimedia Learning}}.
\newblock \bibinfo{publisher}{Cambridge University Press}, \bibinfo{address}{Cambridge}, \bibinfo{pages}{43–71}.
\newblock


\bibitem[McCall(2017)]%
        {McCall2017}
\bibfield{author}{\bibinfo{person}{Cade McCall}.} \bibinfo{year}{2017}\natexlab{}.
\newblock \bibinfo{booktitle}{\emph{Mapping Social Interactions: The Science of Proxemics}}.
\newblock \bibinfo{publisher}{Springer International Publishing}, \bibinfo{address}{Cham}, \bibinfo{pages}{295--308}.
\newblock
\showISBNx{978-3-319-47429-8}
\href{https://doi.org/10.1007/7854_2015_431}{doi:\nolinkurl{10.1007/7854_2015_431}}


\bibitem[Melo et~al\mbox{.}(2023)]%
        {Melo2023}
\bibfield{author}{\bibinfo{person}{Miguel Melo}, \bibinfo{person}{Guilherme Gonçalves}, \bibinfo{person}{josé Vasconcelos-Raposo}, {and} \bibinfo{person}{Maximino Bessa}.} \bibinfo{year}{2023}\natexlab{}.
\newblock \showarticletitle{How Much Presence is Enough? Qualitative Scales for Interpreting the Igroup Presence Questionnaire Score}.
\newblock \bibinfo{journal}{\emph{IEEE Access}}  \bibinfo{volume}{11} (\bibinfo{year}{2023}), \bibinfo{pages}{24675--24685}.
\newblock
\href{https://doi.org/10.1109/ACCESS.2023.3254892}{doi:\nolinkurl{10.1109/ACCESS.2023.3254892}}


\bibitem[Mikhailenko et~al\mbox{.}(2022)]%
        {Mikhailenko2022}
\bibfield{author}{\bibinfo{person}{Maria Mikhailenko}, \bibinfo{person}{Nadezhda Maksimenko}, {and} \bibinfo{person}{Mikhail Kurushkin}.} \bibinfo{year}{2022}\natexlab{}.
\newblock \showarticletitle{Eye-Tracking in Immersive Virtual Reality for Education: A Review of the Current Progress and Applications}.
\newblock \bibinfo{journal}{\emph{Frontiers in Education}}  \bibinfo{volume}{7} (\bibinfo{date}{03} \bibinfo{year}{2022}), \bibinfo{pages}{697032}.
\newblock
\href{https://doi.org/10.3389/feduc.2022.697032}{doi:\nolinkurl{10.3389/feduc.2022.697032}}


\bibitem[Mishra et~al\mbox{.}(2022)]%
        {mishra-etal-2022-reframing}
\bibfield{author}{\bibinfo{person}{Swaroop Mishra}, \bibinfo{person}{Daniel Khashabi}, \bibinfo{person}{Chitta Baral}, \bibinfo{person}{Yejin Choi}, {and} \bibinfo{person}{Hannaneh Hajishirzi}.} \bibinfo{year}{2022}\natexlab{}.
\newblock \showarticletitle{Reframing Instructional Prompts to {GPT}k{'}s Language}. In \bibinfo{booktitle}{\emph{Findings of the Association for Computational Linguistics: ACL 2022}}, \bibfield{editor}{\bibinfo{person}{Smaranda Muresan}, \bibinfo{person}{Preslav Nakov}, {and} \bibinfo{person}{Aline Villavicencio}} (Eds.). \bibinfo{publisher}{Association for Computational Linguistics}, \bibinfo{address}{Dublin, Ireland}, \bibinfo{pages}{589--612}.
\newblock
\href{https://doi.org/10.18653/v1/2022.findings-acl.50}{doi:\nolinkurl{10.18653/v1/2022.findings-acl.50}}


\bibitem[Mohd~Razali and Yap(2011)]%
        {Razali2011}
\bibfield{author}{\bibinfo{person}{Nornadiah Mohd~Razali} {and} \bibinfo{person}{Bee Yap}.} \bibinfo{year}{2011}\natexlab{}.
\newblock \showarticletitle{Power Comparisons of Shapiro-Wilk, Kolmogorov-Smirnov, Lilliefors and Anderson-Darling Tests}.
\newblock \bibinfo{journal}{\emph{J. Stat. Model. Analytics}}  \bibinfo{volume}{2} (\bibinfo{date}{01} \bibinfo{year}{2011}).
\newblock


\bibitem[Moreo et~al\mbox{.}(2022)]%
        {Andrew2022}
\bibfield{author}{\bibinfo{person}{Andrew Moreo}, \bibinfo{person}{Mark Traynor}, {and} \bibinfo{person}{Srikanth Beldona}.} \bibinfo{year}{2022}\natexlab{}.
\newblock \showarticletitle{Food enthusiasts: A behavioral typology}.
\newblock \bibinfo{journal}{\emph{Food Quality and Preference}}  \bibinfo{volume}{96} (\bibinfo{year}{2022}), \bibinfo{pages}{104369}.
\newblock
\showISSN{0950-3293}
\href{https://doi.org/10.1016/j.foodqual.2021.104369}{doi:\nolinkurl{10.1016/j.foodqual.2021.104369}}


\bibitem[Nappi et~al\mbox{.}(2024)]%
        {Nappi2024}
\bibfield{author}{\bibinfo{person}{Maria~Laura Nappi}, \bibinfo{person}{Mario Buono}, \bibinfo{person}{Camelia Chiv{\u{a}}ran}, {and} \bibinfo{person}{Rosa~Maria Giusto}.} \bibinfo{year}{2024}\natexlab{}.
\newblock \showarticletitle{Models and tools for the digital organisation of knowledge: accessible and adaptive narratives for cultural heritage}.
\newblock \bibinfo{journal}{\emph{Heritage Science}} \bibinfo{volume}{12}, \bibinfo{number}{1} (\bibinfo{date}{05 Apr} \bibinfo{year}{2024}), \bibinfo{pages}{112}.
\newblock
\showISSN{2050-7445}
\href{https://doi.org/10.1186/s40494-024-01219-z}{doi:\nolinkurl{10.1186/s40494-024-01219-z}}


\bibitem[Not and Petrelli(2019)]%
        {Not2019}
\bibfield{author}{\bibinfo{person}{Elena Not} {and} \bibinfo{person}{Daniela Petrelli}.} \bibinfo{year}{2019}\natexlab{}.
\newblock \showarticletitle{Empowering cultural heritage professionals with tools for authoring and deploying personalised visitor experiences}.
\newblock \bibinfo{journal}{\emph{User Modeling and User-Adapted Interaction}} \bibinfo{volume}{29}, \bibinfo{number}{1} (\bibinfo{date}{March} \bibinfo{year}{2019}), \bibinfo{pages}{67–120}.
\newblock
\showISSN{0924-1868}
\href{https://doi.org/10.1007/s11257-019-09224-9}{doi:\nolinkurl{10.1007/s11257-019-09224-9}}


\bibitem[Poivet et~al\mbox{.}(2023)]%
        {Poivet2023}
\bibfield{author}{\bibinfo{person}{Remi Poivet}, \bibinfo{person}{Mélanie Lopez~Malet}, \bibinfo{person}{Catherine Pelachaud}, {and} \bibinfo{person}{Malika Auvray}.} \bibinfo{year}{2023}\natexlab{}.
\newblock \showarticletitle{The Influence of Conversational Agents’ Role and Communication Style on User Experience}.
\newblock \bibinfo{journal}{\emph{Frontiers in Psychology}}  \bibinfo{volume}{14} (\bibinfo{year}{2023}), \bibinfo{pages}{1266186}.
\newblock
\showISSN{1664-1078}
\href{https://doi.org/10.3389/fpsyg.2023.1266186}{doi:\nolinkurl{10.3389/fpsyg.2023.1266186}}


\bibitem[Pu and Beam(2024)]%
        {Pu2024}
\bibfield{author}{\bibinfo{person}{Zhixin Pu} {and} \bibinfo{person}{Michael Beam}.} \bibinfo{year}{2024}\natexlab{}.
\newblock \showarticletitle{The impacts of relevance of recommendations and goal commitment on user experience in news recommender design}.
\newblock \bibinfo{journal}{\emph{User Modeling and User-Adapted Interaction}}  \bibinfo{volume}{34} (\bibinfo{date}{06} \bibinfo{year}{2024}), \bibinfo{pages}{925--953}.
\newblock
\href{https://doi.org/10.1007/s11257-024-09405-1}{doi:\nolinkurl{10.1007/s11257-024-09405-1}}


\bibitem[Putze et~al\mbox{.}(2020)]%
        {Putze2020}
\bibfield{author}{\bibinfo{person}{Susanne Putze}, \bibinfo{person}{Dmitry Alexandrovsky}, \bibinfo{person}{Felix Putze}, \bibinfo{person}{Sebastian H\"{o}ffner}, \bibinfo{person}{Jan~David Smeddinck}, {and} \bibinfo{person}{Rainer Malaka}.} \bibinfo{year}{2020}\natexlab{}.
\newblock \showarticletitle{Breaking The Experience: Effects of Questionnaires in VR User Studies}. In \bibinfo{booktitle}{\emph{Proceedings of the 2020 CHI Conference on Human Factors in Computing Systems}} (Honolulu, HI, USA) \emph{(\bibinfo{series}{CHI '20})}. \bibinfo{publisher}{Association for Computing Machinery}, \bibinfo{address}{New York, NY, USA}, \bibinfo{pages}{1–15}.
\newblock
\showISBNx{9781450367080}
\href{https://doi.org/10.1145/3313831.3376144}{doi:\nolinkurl{10.1145/3313831.3376144}}


\bibitem[Python(2021)]%
        {pythonDocumentation}
\bibfield{author}{\bibinfo{person}{Python}.} \bibinfo{year}{2021}\natexlab{}.
\newblock \bibinfo{title}{Python 3.9.6 documentation}.
\newblock
\urldef\tempurl%
\url{https://docs.python.org/release/3.9.6/}
\showURL{%
\tempurl}


\bibitem[Raptis et~al\mbox{.}(2019)]%
        {Raptis2019}
\bibfield{author}{\bibinfo{person}{George~E. Raptis}, \bibinfo{person}{Christos Fidas}, \bibinfo{person}{Christina Katsini}, {and} \bibinfo{person}{Nikolaos Avouris}.} \bibinfo{year}{2019}\natexlab{}.
\newblock \showarticletitle{A cognition-centered personalization framework for cultural-heritage content}.
\newblock \bibinfo{journal}{\emph{User Modeling and User-Adapted Interaction}} \bibinfo{volume}{29}, \bibinfo{number}{1} (\bibinfo{date}{March} \bibinfo{year}{2019}), \bibinfo{pages}{9–65}.
\newblock
\showISSN{0924-1868}
\href{https://doi.org/10.1007/s11257-019-09226-7}{doi:\nolinkurl{10.1007/s11257-019-09226-7}}


\bibitem[Renninger and Hidi(2016)]%
        {Renninger2016}
\bibfield{author}{\bibinfo{person}{K. Renninger} {and} \bibinfo{person}{Suzanne Hidi}.} \bibinfo{year}{2016}\natexlab{}.
\newblock \bibinfo{booktitle}{\emph{The Power of Interest for Motivation and Engagement}}.
\newblock \bibinfo{publisher}{Routledge}, \bibinfo{address}{New York, NY, USA}. 1--177 pages.
\newblock
\showISBNx{9781315771045}
\href{https://doi.org/10.4324/9781315771045}{doi:\nolinkurl{10.4324/9781315771045}}


\bibitem[Sahoo et~al\mbox{.}(2024)]%
        {sahoo2024systematicsurveypromptengineering}
\bibfield{author}{\bibinfo{person}{Pranab Sahoo}, \bibinfo{person}{Ayush~Kumar Singh}, \bibinfo{person}{Sriparna Saha}, \bibinfo{person}{Vinija Jain}, \bibinfo{person}{Samrat Mondal}, {and} \bibinfo{person}{Aman Chadha}.} \bibinfo{year}{2024}\natexlab{}.
\newblock \bibinfo{title}{A Systematic Survey of Prompt Engineering in Large Language Models: Techniques and Applications}.
\newblock
\showeprint[arxiv]{2402.07927}~[cs.AI]
\urldef\tempurl%
\url{https://arxiv.org/abs/2402.07927}
\showURL{%
\tempurl}


\bibitem[Schilbach(2015)]%
        {SCHILBACH2015130}
\bibfield{author}{\bibinfo{person}{Leonhard Schilbach}.} \bibinfo{year}{2015}\natexlab{}.
\newblock \showarticletitle{Eye to eye, face to face and brain to brain: novel approaches to study the behavioral dynamics and neural mechanisms of social interactions}.
\newblock \bibinfo{journal}{\emph{Current Opinion in Behavioral Sciences}}  \bibinfo{volume}{3} (\bibinfo{year}{2015}), \bibinfo{pages}{130--135}.
\newblock
\showISSN{2352-1546}
\href{https://doi.org/10.1016/j.cobeha.2015.03.006}{doi:\nolinkurl{10.1016/j.cobeha.2015.03.006}}
\newblock
\shownote{Social behavior}.


\bibitem[SciPy(2024)]%
        {SciPy}
\bibfield{author}{\bibinfo{person}{SciPy}.} \bibinfo{year}{2024}\natexlab{}.
\newblock \bibinfo{title}{SciPy 1.13.1 documentation}.
\newblock
\urldef\tempurl%
\url{https://docs.scipy.org/doc/scipy-1.13.1/}
\showURL{%
\tempurl}


\bibitem[Soler-Dominguez et~al\mbox{.}(2017)]%
        {Dominguez2017}
\bibfield{author}{\bibinfo{person}{Jose~L. Soler-Dominguez}, \bibinfo{person}{Jorge~D. Camba}, \bibinfo{person}{Manuel Contero}, {and} \bibinfo{person}{Mariano Alca{\~{n}}iz}.} \bibinfo{year}{2017}\natexlab{}.
\newblock \showarticletitle{A Proposal for the Selection of Eye-Tracking Metrics for the Implementation of Adaptive Gameplay in Virtual Reality Based Games}. In \bibinfo{booktitle}{\emph{Virtual, Augmented and Mixed Reality}}, \bibfield{editor}{\bibinfo{person}{Stephanie Lackey} {and} \bibinfo{person}{Jessie Chen}} (Eds.). \bibinfo{publisher}{Springer International Publishing}, \bibinfo{address}{Cham}, \bibinfo{pages}{369--380}.
\newblock
\showISBNx{978-3-319-57987-0}


\bibitem[Stazio(2021)]%
        {Stazio2021}
\bibfield{author}{\bibinfo{person}{Marialuisa Stazio}.} \bibinfo{year}{2021}\natexlab{}.
\newblock \showarticletitle{Verace Glocal Pizza. Localized globalism and globalized localism in the Neapolitan artisan pizza}.
\newblock \bibinfo{journal}{\emph{Food, Culture \& Society}}  \bibinfo{volume}{24} (\bibinfo{date}{03} \bibinfo{year}{2021}), \bibinfo{pages}{1--25}.
\newblock
\href{https://doi.org/10.1080/15528014.2021.1884400}{doi:\nolinkurl{10.1080/15528014.2021.1884400}}


\bibitem[Stock et~al\mbox{.}(2007)]%
        {Stock2007}
\bibfield{author}{\bibinfo{person}{Oliviero Stock}, \bibinfo{person}{Massimo Zancanaro}, \bibinfo{person}{Paolo Busetta}, \bibinfo{person}{Charles Callaway}, \bibinfo{person}{Antonio Krüger}, \bibinfo{person}{Michael Kruppa}, \bibinfo{person}{Tsvi Kuflik}, \bibinfo{person}{Elena Not}, {and} \bibinfo{person}{Cesare Rocchi}.} \bibinfo{year}{2007}\natexlab{}.
\newblock \showarticletitle{Adaptive, intelligent presentation of information for the museum visitor in PEACH}.
\newblock \bibinfo{journal}{\emph{User Model. User-Adapt. Interact.}}  \bibinfo{volume}{17} (\bibinfo{date}{05} \bibinfo{year}{2007}), \bibinfo{pages}{257--304}.
\newblock
\href{https://doi.org/10.1007/s11257-007-9029-6}{doi:\nolinkurl{10.1007/s11257-007-9029-6}}


\bibitem[Sweller et~al\mbox{.}(2011)]%
        {Sweller2011}
\bibfield{author}{\bibinfo{person}{John Sweller}, \bibinfo{person}{Paul Ayres}, {and} \bibinfo{person}{Slava Kalyuga}.} \bibinfo{year}{2011}\natexlab{}.
\newblock \bibinfo{booktitle}{\emph{The Guidance Fading Effect}}.
\newblock \bibinfo{publisher}{Springer New York}, \bibinfo{address}{New York, NY}, \bibinfo{pages}{171--182}.
\newblock
\href{https://doi.org/10.1007/978-1-4419-8126-4_13}{doi:\nolinkurl{10.1007/978-1-4419-8126-4_13}}


\bibitem[Trubek et~al\mbox{.}(2017)]%
        {TRUBEK2017297}
\bibfield{author}{\bibinfo{person}{Amy~B. Trubek}, \bibinfo{person}{Maria Carabello}, \bibinfo{person}{Caitlin Morgan}, {and} \bibinfo{person}{Jacob Lahne}.} \bibinfo{year}{2017}\natexlab{}.
\newblock \showarticletitle{Empowered to cook: The crucial role of ‘food agency’ in making meals}.
\newblock \bibinfo{journal}{\emph{Appetite}}  \bibinfo{volume}{116} (\bibinfo{year}{2017}), \bibinfo{pages}{297--305}.
\newblock
\showISSN{0195-6663}
\href{https://doi.org/10.1016/j.appet.2017.05.017}{doi:\nolinkurl{10.1016/j.appet.2017.05.017}}


\bibitem[UNESCO(2004)]%
        {unesco2004yamato}
\bibfield{author}{\bibinfo{person}{UNESCO}.} \bibinfo{year}{2004}\natexlab{}.
\newblock \bibinfo{title}{Co-operation and coordination between the UNESCO Conventions concerning heritage: the Yamato Declaration on Integrated Approaches for Safeguarding Tangible and Intangible Cultural Heritage}.
\newblock
\newblock
\shownote{World Heritage Committee, 7th extraordinary session, Paris, 2004. Document code: WHC.2004/CONF.202/CLD.21, WHC-04/7 EXT.COM/INF.9}.


\bibitem[UNESCO(2024)]%
        {unesco2024oral}
\bibfield{author}{\bibinfo{person}{UNESCO}.} \bibinfo{year}{2024}\natexlab{}.
\newblock \bibinfo{title}{Oral traditions and expressions including language as a vehicle of the intangible cultural heritage}.
\newblock
\urldef\tempurl%
\url{https://ich.unesco.org/en/oral-traditions-and-expressions-00053}
\showURL{%
\tempurl}
\newblock
\shownote{Retrieved July 10, 2024}.


\bibitem[{United Nations Educational, Scientific and Cultural Organization}(2017)]%
        {UNESCO_Pizza_Inscription}
\bibfield{author}{\bibinfo{person}{{United Nations Educational, Scientific and Cultural Organization}}.} \bibinfo{year}{2017}\natexlab{}.
\newblock \bibinfo{booktitle}{\emph{Art of Neapolitan ‘Pizzaiuolo’}}.
\newblock UNESCO.
\newblock
\urldef\tempurl%
\url{https://ich.unesco.org/en/RL/art-of-neapolitan-pizzaiuolo-00722}
\showURL{%
\tempurl}
\newblock
\shownote{Accessed: July 17, 2024}.


\bibitem[Vallat(2024)]%
        {Pingouin}
\bibfield{author}{\bibinfo{person}{Raphael Vallat}.} \bibinfo{year}{2018-2024}\natexlab{}.
\newblock \bibinfo{title}{Pingouin documentation}.
\newblock
\urldef\tempurl%
\url{https://pingouin-stats.org/build/html/index.html}
\showURL{%
\tempurl}


\bibitem[Walkington and Bernacki(2014)]%
        {Walkington2014}
\bibfield{author}{\bibinfo{person}{Candace Walkington} {and} \bibinfo{person}{Matthew Bernacki}.} \bibinfo{year}{2014}\natexlab{}.
\newblock \bibinfo{booktitle}{\emph{Motivating Students by “Personalizing” Learning around Individual Interests: A Consideration of Theory, Design, and Implementation Issues}}. Vol.~\bibinfo{volume}{18}.
\newblock \bibinfo{publisher}{Emerald Group Publishing Limited}, \bibinfo{address}{Bingley, UK}, \bibinfo{pages}{139--176}.
\newblock
\showISBNx{978-1-78350-555-5}
\href{https://doi.org/10.1108/S0749-742320140000018004}{doi:\nolinkurl{10.1108/S0749-742320140000018004}}


\bibitem[Wang et~al\mbox{.}(2024)]%
        {wangjiayin2024}
\bibfield{author}{\bibinfo{person}{Jiayin Wang}, \bibinfo{person}{Weizhi Ma}, \bibinfo{person}{Peijie Sun}, \bibinfo{person}{Min Zhang}, {and} \bibinfo{person}{Jian-Yun Nie}.} \bibinfo{year}{2024}\natexlab{}.
\newblock \bibinfo{title}{Understanding User Experience in Large Language Model Interactions}.
\newblock
\showeprint[arxiv]{2401.08329}~[cs.HC]
\urldef\tempurl%
\url{https://arxiv.org/abs/2401.08329}
\showURL{%
\tempurl}


\bibitem[Weil(1999)]%
        {weil2007being}
\bibfield{author}{\bibinfo{person}{Stephen~E. Weil}.} \bibinfo{year}{1999}\natexlab{}.
\newblock \showarticletitle{From Being about Something to Being for Somebody: The Ongoing Transformation of the American Museum}.
\newblock \bibinfo{journal}{\emph{Daedalus}} \bibinfo{volume}{128}, \bibinfo{number}{3} (\bibinfo{year}{1999}), \bibinfo{pages}{229--258}.
\newblock
\showISSN{00115266}
\urldef\tempurl%
\url{http://www.jstor.org/stable/20027573}
\showURL{%
\tempurl}


\bibitem[Wilk(2006)]%
        {wilk2006fast}
\bibfield{author}{\bibinfo{person}{Richard Wilk}.} \bibinfo{year}{2006}\natexlab{}.
\newblock \bibinfo{booktitle}{\emph{Fast Food/Slow Food: The Cultural Economy of the Global Food System}}.
\newblock \bibinfo{publisher}{AltaMira Press}, \bibinfo{address}{Lanham, MD, USA}.
\newblock


\bibitem[Xu et~al\mbox{.}(2023)]%
        {Xu2023}
\bibfield{author}{\bibinfo{person}{Wei Xu}, \bibinfo{person}{Marvin Dainoff}, \bibinfo{person}{Liezhong Ge}, {and} \bibinfo{person}{Zaifeng Gao}.} \bibinfo{year}{2023}\natexlab{}.
\newblock \showarticletitle{Transitioning to Human Interaction with AI Systems: New Challenges and Opportunities for HCI Professionals to Enable Human-Centered AI}.
\newblock \bibinfo{journal}{\emph{International Journal of Human-Computer Interaction}}  \bibinfo{volume}{39} (\bibinfo{date}{01} \bibinfo{year}{2023}), \bibinfo{pages}{494--518}.
\newblock
\href{https://doi.org/10.1080/10447318.2022.2041900}{doi:\nolinkurl{10.1080/10447318.2022.2041900}}


\bibitem[Yaremych and Persky(2019)]%
        {YAREMYCH2019103845}
\bibfield{author}{\bibinfo{person}{Haley~E. Yaremych} {and} \bibinfo{person}{Susan Persky}.} \bibinfo{year}{2019}\natexlab{}.
\newblock \showarticletitle{Tracing physical behavior in virtual reality: A narrative review of applications to social psychology}.
\newblock \bibinfo{journal}{\emph{Journal of Experimental Social Psychology}}  \bibinfo{volume}{85} (\bibinfo{year}{2019}), \bibinfo{pages}{103845}.
\newblock
\showISSN{0022-1031}
\href{https://doi.org/10.1016/j.jesp.2019.103845}{doi:\nolinkurl{10.1016/j.jesp.2019.103845}}


\bibitem[Zander et~al\mbox{.}(2015)]%
        {Zander2015}
\bibfield{author}{\bibinfo{person}{Steffi Zander}, \bibinfo{person}{Maria Reichelt}, \bibinfo{person}{Stefanie Wetzel}, \bibinfo{person}{Sven Bertel}, {and} \bibinfo{person}{Frauke Kämmerer}.} \bibinfo{year}{2015}\natexlab{}.
\newblock \showarticletitle{Does Personalisation Promote Learners' Attention? An Eye-Tracking Study}.
\newblock \bibinfo{journal}{\emph{Frontline Learning Research}}  \bibinfo{volume}{3} (\bibinfo{date}{11} \bibinfo{year}{2015}), \bibinfo{pages}{1--13}.
\newblock
\href{https://doi.org/10.14786/flr.v3i4.161}{doi:\nolinkurl{10.14786/flr.v3i4.161}}


\bibitem[Zhang et~al\mbox{.}(2024)]%
        {zhang2024}
\bibfield{author}{\bibinfo{person}{Lichao Zhang}, \bibinfo{person}{Jia Yu}, \bibinfo{person}{Shuai Zhang}, \bibinfo{person}{Long Li}, \bibinfo{person}{Yangyang Zhong}, \bibinfo{person}{Guanbao Liang}, \bibinfo{person}{Yuming Yan}, \bibinfo{person}{Qing Ma}, \bibinfo{person}{Fangsheng Weng}, \bibinfo{person}{Fayu Pan}, \bibinfo{person}{Jing Li}, \bibinfo{person}{Renjun Xu}, {and} \bibinfo{person}{Zhenzhong Lan}.} \bibinfo{year}{2024}\natexlab{}.
\newblock \bibinfo{title}{Unveiling the Impact of Multi-Modal Interactions on User Engagement: A Comprehensive Evaluation in AI-driven Conversations}.
\newblock
\showeprint[arxiv]{2406.15000}~[cs.CL]
\urldef\tempurl%
\url{https://arxiv.org/abs/2406.15000}
\showURL{%
\tempurl}


\end{thebibliography}

\appendix
\clearpage

\section{Adaptive Guidance Dialogue}

\begin{table}[H]
\caption{Example of after ingredient selection instructions in each adaptivity condition.}
\label{tab:tutor_scripts}
\centering
\begin{tabular}{@{}p{0.2\columnwidth} p{0.75\columnwidth}@{}}
\toprule
\textbf{Condition} & \textbf{Tutor instruction (excerpt)} \\ \midrule
None &
``Great choice! Let's get started on the dough. Combine flour, water, yeast, and salt. Neapolitan pizza is all about simplicity and authenticity.'' \\ 
\midrule
Moderate &
``Perfect, you've selected the core ingredients for Neapolitan dough. Start with Type 00 flour, it's finely milled for a soft crust. Mix flour and salt, then make a well for the yeast-water mixture. Neapolitans often let the dough rise slowly for better texture. You're on your way to a classic base.'' \\
\midrule
High &
``Magnificent! In Naples, pizza starts with just flour, water, yeast, and salt. Try dissolving the yeast in warm water for even mixing. Type 00 flour is traditional, but you're free to explore. Begin kneading gently. Neapolitans believe that patient hands give dough its soul. Embrace the craft.'' \\

\bottomrule
\end{tabular}
\end{table}

\section{Post–assessment Interest in Cultural Heritage Activities}
\begin{figure}[ht]
\centering
\includegraphics[width=0.7\columnwidth]{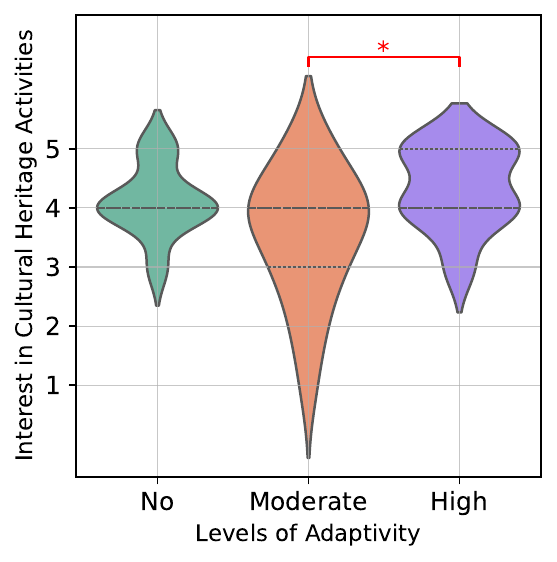}
\caption{Responses to the post-study survey question: \textit{``After the user study, would you be more interested in participating in cultural heritage activities such as visiting museums, participating in local workshops, or volunteering in cultural events?''} (1: ``Definitely not'' to 5: ``Definitely would'') across adaptivity conditions (No, Moderate, High). Significance level indicated by *, corresponding to \(p < .05\) (Kruskal–Wallis \(H(2)=6.226,\ p=0.044\); Dunn’s post-hoc).}
\Description{Responses to the post-study survey question: \textit{``After the user study, would you be more interested in participating in cultural heritage activities such as visiting museums, participating in local workshops, or volunteering in cultural events?''} (1: ``Definitely not'' to 5: ``Definitely would'') across adaptivity conditions (No, Moderate, High). Significance level indicated by *, corresponding to \(p < .05\) (Kruskal–Wallis \(H(2)=6.226,\ p=0.044\); Dunn’s post-hoc).}
\label{fig:interest_violin}
\end{figure}

\newpage
\section{Pre-assessment Questionnaire}

\begin{table}[htp!]
\caption{Pre-assessment items measuring participants' engagement with cultural heritage activities, preferred learning methods, and motivations. Cultural engagement and motivation items were rated individually on a 5-point Likert scale (1 = Not at all motivated, 5 = Very strongly motivated), while learning preferences were assessed through multiple-choice selection.}

\label{tab:heritage}
\begin{tabular}{@{}p{0.55\columnwidth} p{0.4\columnwidth}@{}}
\toprule
\textbf{Question} & \textbf{Scale/Type} \\
\midrule
Participation in cultural heritage activities & 1 (Never) -- 5 (Very Frequently) \\
Preferred learning methods (books, sites, online, etc.) & Multiple choice \\
Motivation: Personal/family history & 1 (Not) -- 5 (Strong) \\
Motivation: Educational purposes & 1 (Not) -- 5 (Strong) \\
Motivation: Entertainment & 1 (Not) -- 5 (Strong) \\
Motivation: Socializing/friends & 1 (Not) -- 5 (Strong) \\
\bottomrule
\end{tabular}
\end{table}

\begin{table}[htp!]
\caption{Pre-assessment items evaluating participants' interest levels across various cultural heritage domains.}
\label{tab:interest}
\begin{tabular}{@{}p{0.45\columnwidth} p{0.5\columnwidth}@{}}
\toprule
\textbf{Domain} & \textbf{Interest Scale} \\
\midrule
Culinary arts & 1 (Not at all) -- 5 (Extremely) \\
History and archaeology & 1 -- 5 \\
Arts and crafts & 1 -- 5 \\
Traditional music and dance & 1 -- 5 \\
Linguistics and storytelling & 1 -- 5 \\
\bottomrule
\end{tabular}
\end{table}

\begin{table}[htp!]
\caption{Pre-assessment items measuring VR experience and baseline pizza-making knowledge.}
\label{tab:vrpizza}
\centering
\begin{tabular}{@{}p{0.65\columnwidth} p{0.28\columnwidth}@{}}
\toprule
\textbf{Item} & \textbf{Response Type} \\
\midrule
Used VR before & Yes/No \\
VR usage frequency & 5-point \\
VR content experienced (games, education, simulation, etc.) & Multiple choice \\
Key ingredients of Neapolitan pizza & Open-ended \\
Steps in making pizza & Open-ended \\
Cooking technique explanation & Open-ended \\
\bottomrule
\end{tabular}
\end{table}

\begin{table*}[htbp]
  \centering
  \caption{Pre-assessment items measuring participants' culinary creativity, enthusiasm, and attitudes toward cooking. All items were rated on a 5-point Likert scale (1 = Strongly Disagree, 5 = Strongly Agree).}
  \label{tab:culinary}
  \begin{tabular}{@{}p{0.7\textwidth} c@{}}
    \toprule
    \textbf{Statement} & \textbf{Agreement Scale} \\
    \midrule
    I actively seek out food-related experiences and knowledge. & 1--5 \\
    I have taken non-professional cooking classes to improve my culinary skills. & 1--5 \\
    I enjoy incorporating unique or authentic food experiences into my travel. & 1--5 \\
    Travelling to experience local cuisines is important to me. & 1--5 \\
    I frequently explore new and different food experiences, cuisines, and flavours. & 1--5 \\
    Trying out newly opened restaurants excites me. & 1--5 \\
    I enjoy experimenting with flavour combinations in my cooking. & 1--5 \\
    Watching television shows about cooking techniques and ingredients is a regular activity for me. & 1--5 \\
    I am keen on cooking competition shows and programs about the cultural aspects of food. & 1--5 \\
    Cooking from scratch at home is a practice I engage in regularly. & 1--5 \\
    I am always on the lookout for new recipes to try at home. & 1--5 \\
    \bottomrule
  \end{tabular}
\end{table*}

\end{document}